\documentclass[fleqn,usenatbib]{mnras}
\usepackage{newtxtext,newtxmath}

\usepackage[T1]{fontenc}

\DeclareRobustCommand{\VAN}[3]{#2}
\let\VANthebibliography\thebibliography
\def\thebibliography{\DeclareRobustCommand{\VAN}[3]{##3}\VANthebibliography}

\usepackage{graphicx}	
\usepackage{amsmath}	

\usepackage{amssymb}
\usepackage{gensymb}
\usepackage[nopatch]{microtype}
\usepackage{booktabs}
\usepackage{verbatim}
\usepackage{orcidlink}
\usepackage{lipsum}

\DeclareUnicodeCharacter{2212}{-}

\newcommand{\kms}{~km~s$^{-1}$}
\newcommand{\baygaud}{\texttt{BAYGAUD-PI}}
\newcommand{\Hone}{H{\sc i}}
\newcommand{\Htwo}{H$_2$}

\newcommand{\Mstar}{$M_{\rm \star}$}
\newcommand{\Msol}{$M_{\rm \odot}$}
\newcommand{\Zsol}{$Z_{\rm \odot}$}

\newcommand{\logOH}{12 + log(O/H)}

\newcommand{\Msolyr}{$M_{\rm \odot}$yr$^{-1}$}
\newcommand{\fn}{$f_{\rm n}$}

\newcommand{\medianfn}{$\langle$\fn$\rangle$}

\title[Colder HI in nearby galaxies]{Exploring the potential for kinematically colder \Hone\ component as a tracer for star-forming gas in nearby galaxies}

\author[H.-J. Park et al.]{
Hye-Jin Park$^{\orcidlink{0000-0002-9809-6631}}$$^{1, 2}$\thanks{hyejin.park@anu.edu.au}
Andrew J. Battisti$^{\orcidlink{0000-0003-4569-2285}}$$^{1, 3, 2}$
Antoine Marchal$^{\orcidlink{https://orcid.org/0000-0002-5501-232X}}$$^{1}$
Luca Cortese$^{\orcidlink{0000-0002-7422-9823}}$$^{3, 2}$
Emily Wisnioski$^{\orcidlink{0000-0003-1657-7878}}$$^{1, 2}$
\newauthor
Mark Seibert$^{\orcidlink{0000-0002-1143-5515}}$$^{4}$
Shin-Jeong Kim$^{\orcidlink{0000-0002-6760-7531}}$$^{1}$
Naomi McClure-Griffiths$^{\orcidlink{0000-0003-2730-
957X}}$$^{1}$
W.J.G. de Blok$^{\orcidlink{0000-0001-8957-4518}}$,$^{5,6,7}$
\newauthor
Kathryn Grasha$^{\orcidlink{0000-0002-3247-5321}}$$^{1, 2}$
Barry F. Madore$^{\orcidlink{0000-0002-1576-1676}}$,$^{4,8}$
Jeff A. Rich$^{\orcidlink{0000-0002-5807-5078}}$, $^{4}$
Rachael L. Beaton$^{\orcidlink{0000-0002-1691-8217}}$, $^{4,9}$
\\
$^{1}$Research School of Astronomy and Astrophysics, Australian National University, Cotter Road, Weston Creek, ACT 2611, Australia\\
$^{2}$ARC Centre of Excellence for All Sky Astrophysics in 3 Dimensions (ASTRO 3D), Australia\\
$^{3}$International Centre for Radio Astronomy Research (ICRAR), University of Western Australia, M468, 35 Stirling Highway, Crawley, WA 6009, Australia\\
$^{4}$The Observatories, Carnegie Institution for Science, 813 Santa Barbara Street, Pasadena, CA 91101, USA\\
$^{5}$ASTRON, the Netherlands Institute for Radio Astronomy, Oude Hoogeveensedijk 4, 7991 PD Dwingeloo, the Netherlands\\
$^{6}$Department of Astronomy, University of Cape Town, Private Bag X3, Rondebosch 7701, South Africa\\
$^{7}$Kapteyn Astronomical Institute, University of Groningen, PO Box 800, 9700 AV Groningen, The Netherlands\\
$^{8}$Department of Astronomy and Astrophysics, University of Chicago, Chicago, IL, USA\\
$^{9}$Department of Astrophysical Sciences, 4 Ivy Lane, Princeton University, Princeton, NJ 08544, USA\\
}

\date{Accepted XXX. Received YYY; in original form ZZZ}

\pubyear{\the\year{}}

\begin{document}
\label{firstpage}
\pagerange{\pageref{firstpage}--\pageref{lastpage}}
\maketitle


\begin{abstract}
Atomic hydrogen (\Hone) dominates the mass of the cold interstellar medium, undergoing thermal condensation to form molecular gas and fuel star formation. Kinematically colder \Hone\ components, identified via kinematic decomposition of \Hone\ 21~cm data cubes, serve as a crucial transition phase between diffuse warm neutral gas and molecular hydrogen (\Htwo). We analyse these colder \Hone\ components by decomposing \Hone\ 21~cm data cubes of seven nearby galaxies -- Sextans~A, NGC~6822, WLM, NGC~5068, NGC~7793, NGC~1566, and NGC~5236 -- spanning metallicities (0.1 $<$ Z/\Zsol $<$ 1.0) and physical scales (53\--1134~pc). Using a velocity dispersion threshold of 6~\kms, we classify the kinematically distinct components into narrow (colder) and broad (warmer). Cross-correlation analysis between the narrow \Hone\ components and \Htwo\ or star formation rate (SFR) surface density at different spatial scales reveals that dwarf galaxies exhibit the strongest correlation at $\sim$500-700~pc. The radially binned narrow \Hone\ fraction, \fn=$I_{\rm narrow HI} / I_{\rm total HI}$, in dwarf galaxies shows no clear trend with metallicity or SFR, while in spirals, \fn\ is lower in inner regions with higher metallicity and SFR. We find that the dataset resolution significantly impacts the results, with higher physical resolution data yielding a higher median \fn, \medianfn, per galaxy. With this considered, dwarf galaxies consistently exhibit a larger \fn\ than spiral galaxies. These findings highlight the critical role of cold \Hone\ in regulating star formation across different galactic environments and emphasise the need for high-resolution \Hone\ observations to further unravel the connection between atomic-to-molecular gas conversion and galaxy evolution.
\end{abstract}




\section{Introduction}
\label{sec:introduction}
Molecular gas in the interstellar medium (ISM) is an important ingredient for star formation in a galaxy and, consequently, galaxy evolution \citep[][]{tacconi2020evolution, Schinnerer2024molecular}. The molecular gas phase is cold (kinetic temperature, $T_{\rm k}\sim$~10--50~K), and dense ($n_{\rm H}\sim$~10$^{3}$--10$^{6}$~cm$^{-3}$), containing predominantly molecular hydrogen (\Htwo), C, and other molecules such as CO (second most abundant molecule) \citep[see Table 1.3 in ][]{draine2011physics}.


While \Htwo\ is challenging to observe directly due to its lack of a permanent dipole moment, the emission lines of CO in the (sub-)mm wavelengths are typically much easier to observe and are a good proxy for the presence of \Htwo\ and its abundance \citep[for review, see][]{bolatto2013co}. The CO~(1-0) (or CO~(2-1)) observations using (sub-)mm wavelength radio telescopes (e.g., MOPRA-22m, IRAM-30m, and ALMA interferometer) have been used to estimate the total molecular gas mass in a system, by assuming a CO-to-H$_2$ conversion factor, $X_{\rm CO}$. However, the situation significantly changes for low-metallicity galaxies as the CO line observations become challenging due to their lower abundance due to the lower abundance of metals itself, and their molecular gas reservoir is often referred to as being `CO-dark' \citep[e.g.,][]{bolatto2013co, madden2020tracing}. More specifically, the more effective penetration of high-energy photons through the low-metal/dust environment can consequently lead to the photodissociation of CO molecules. As a result, the CO-to-H$_2$ conversion factor is often scaled by metallicity to account for the amount of CO-dark gas \citep[e.g., ][]{narayanan2012general, amorin2016molecular}. However, this conversion is still highly uncertain (\citealt{Schinnerer2024molecular}), prohibiting us from optimally tracing molecular gas components in low metallicity systems, which often occurs for low-mass galaxies in the local Universe but is also important for galaxies in the early Universe.


The observational difficulties in observing CO lines in low-metallicity systems lead to the question of \textit{How can we trace star-forming gas in low-metallicity systems?} Understanding this is crucial, as galaxies in the early universe likely formed in low-metallicity conditions, where the baryonic mass was dominated by the atomic phase and star formation took place within it. In \citet{dickey2000cold}, they found that the cold \Hone\ components in the Small Magellanic Clouds (SMC; $\sim$0.2~\Zsol) have the kinetic temperature of $T_{\rm k}\sim$10--40~K, the temperature range of molecular clouds at Milky Way (MW) metallicity. Moreover, simulations \citep[e.g.,][]{hu2016star, hu2022dependence, seifried2022on} indicate that in low dust/metal systems, \Htwo\ may be quickly destroyed by the radiation from young stars before it has the chance or opacity to fully form, in addition to the early photodissociation of CO molecules. As a result, much of the surrounding gas that can fuel star formation could remain as cold \Hone.

Cold \Hone, or Cold Neutral Medium (CNM; $T_{\rm k} \sim$ 25--250~K; $\rm n_{H} \sim$ 5--20~cm$^{-3}$ in MW) has been studied via \Hone\ 21~cm absorption-emission joint observations against background radio sources \citep[e.g.,][]{murray2018sponge, jameson2019atca, allison2022first, mccluregriffiths2023atomic}. This has primarily been undertaken in very nearby sources such as the MW \citep[e.g.,][]{murray2018sponge} and Magellanic Clouds \citep[e.g.,][]{jameson2019atca, dempsey2022gaskap} where large numbers of background sources are available (due to the large angular size of these foreground sources). However, understanding CNM through the combination of 21~cm absorption and emission lines becomes a non-trivial task for farther systems, i.e., external galaxies, due to the limited bright background source density \citep[$\gtrsim$15~mJy/beam; e.g.,][]{pingel2024local} and limited resolution both in spectral and spatial \citep[][]{koch2021lack}. As a reference, the recent Local Group L-Band Survey (LGLBS\footnote{\url{https://www.lglbs.org/home}}) on NGC~6822 \citep[][]{pingel2024local}, which is one of the nearest external galaxies (D $\sim$~0.5~Mpc) following the Magellanic Clouds, detected only two sightlines having 21~cm absorption features with the emission line available simultaneously.

An alternative way to trace cold \Hone\ components in external galaxies is to utilise \Hone\ 21~cm emission lines. \Hone\ gas is sensitive to diverse galactic processes such as gas accretion, interactions with other galaxies, and outflows. Additionally, the \Hone\ emission line can be a mixture of warm neutral medium (WNM, $\rm n_{H}\sim$~0.03--1.3~cm$^{-3}$; $T_{\rm s}\sim$~4000--8000~K) of a wide velocity dispersion ($\sigma_{\rm V}$ of 5--8\kms) with superimposed CNM \citep[$\sigma_{\rm V} \sim$ 0.4--1\kms; for a review, see][]{mccluregriffiths2023atomic, hunter2024interstellar}, and internal motions within the gas disk (i.e., turbulence). These often cause the \Hone\ emission line velocity profiles to be complex and non-Gaussian.

Over the past two decades, kinematic decomposition techniques have been widely used to disentangle the mixture of the dynamical sources and to trace cold \Hone\ gas in nearby galaxies \citep[e.g.,][]{braun1997temperature, young2003star, deblok2006star, warren2012tracing, saikia2020gas, park2022gas, oh2022kinematic}. These studies isolate \Hone\ components with narrow velocity dispersion ($\sigma_{\rm V} <$ 4--6\kms) and define them as kinematically cold. It has been found that the kinematically cold components are often clumpier than those with broader velocity dispersion and tend to link to regions of active star formation morphologically.

However, previous studies that identified colder or narrow \Hone\ components through kinematic decomposition often focused on either high- or low-metallicity galaxies or individual galaxies, analysing them separately with different criteria to define colder \Hone. For example, \citet{warren2012tracing} studied 31 dwarf galaxies using \Hone\ data from the VLA-ANGST \citep[][]{ott2012vlaangst} and THINGS \citep[][]{walter2008things} surveys, identifying colder \Hone\ components with low-velocity dispersion ($\sigma_{\rm V} <$ 6\kms) through double Gaussian decomposition. Similarly, \citet{braun1997temperature} investigated 11 nearby spiral galaxies and isolated colder \Hone\ components (Full-width-of-half-maximum, FWHM$_{\rm V} <$ 6\kms)\footnote{FWHM $\simeq$ 2.355 $\times \sigma$} and found that such components are clumpy and preferentially located along the spiral arms. Additionally, \citet{park2022gas} detected colder \Hone\ components ($\sigma_{\rm V} <$ 4\kms) in the nearby dwarf galaxy NGC~6822, employing an optimised Gaussian fitting method that allowed up to three components per line-of-sight. To explore colder \Hone\ components across multiple galaxies and the impact of metal abundance, it is necessary to apply a consistent method for distinguishing kinematically distinct components for multiple galaxies, across a wide range of metallicity.

In this study, as a first step towards tracking the cold \Hone\ components, we use a kinematic decomposition tool on the 21~cm \Hone\ data and investigate colder, narrow \Hone\ components in nearby galaxies spanning a range of metallicities (0.1 $<$ Z/\Zsol $<$ 1.0). Specifically, we examine the surface density and mass fraction of the narrow \Hone\ components and analyse their correlation with molecular gas surface density, traced by CO (2-1) or CO (1-0) line observations. Additionally, we explore their relationship with other ISM properties, including star formation rates (SFRs) derived from UV+IR and gas-phase metallicity from optical emission lines.

Mapping gas-phase metallicity from the optical emission line diagnostics often requires Integral Field Spectroscopy (IFS) data. In this work, we utilise the TYPHOON survey\footnote{\url{https://typhoon.datacentral.org.au}} (Carnegie Observatories, TYPHOON Programme PI: Barry F. Madore), a pseudo-IFS survey of 44 nearby galaxies conducted with the du Pont 2.5m telescope at Las Campanas Observatory, Chile. The survey's large field of view (FoV), constructed using a step-and-stare method with a long slit, makes it uniquely suited for studying the ISM, encompassing the outskirts of nearby galaxies. This is particularly relevant given that \Hone\ gas disks are observed to extend significantly farther—up to several times larger—than stellar or molecular gas disks.

This paper is structured as follows: In Sec.~\ref{sec:sample_data}, we introduce our sample galaxies, their characteristics, and the data used in this study, including TYPHOON optical IFS, CO, multi-wavelength, and \Hone\ data. In Sec.~\ref{sec:parameter}, we describe the derivation of key parameters such as \Hone\ gas mass, molecular gas mass, SFR, and gas-phase metallicity. Sec.~\ref{sec:hi_decomposition} details the Gaussian decomposition tool \baygaud\ and our classification of narrow (colder) \Hone\ components. In Sec.~\ref{sec:analysis}, we present our morphological analysis, including 2D cross-correlation studies. Sec.~\ref{sec:discussion} discusses radially binned \fn\ trends with metallicity and SFR comparisons, and also covers the limitations of our study. Finally, we summarise and conclude our work in Sec.~\ref{sec:summary}.

\section{Sample and Data}
\label{sec:sample_data}

\subsection{Sample galaxies}




\begin{table*} 
    \centering
    \begin{tabular}{c|c|c|c|c|c|c|c|c|c|c|c}
    \hline
    Name     & RA & DEC & Dist. & $r_{25}$ & P.A. & Incl. & T-Type & log\Mstar & logSFR$^{\dagger}$ & 12+log(O/H)$^{\dagger}$ & A(V)$_{\rm MW}$$^\ast$ \\
        & J2000 & J2000 & Mpc & arcsec & deg & deg & & $M_{\odot}$ & $M_{\odot}$ yr$^{-1}$ & & mag \\
    \hline
    Sextans~A & 10 11 00.80 & -04 41 34.0 & 1.4 & 176.7 & 86$^{\rm N18}$ & 34$^{\rm N18}$ & 9.5 & 8.04$^{\rm LE19}$ & -2.32 & 7.53 & 0.1198 \\
    NGC~6822 & 19 44 57.74 & -14 48 12.4 & 0.5 & 464.7 & 118$^{\rm N17}$ & 66$^{\rm N17}$ & 9.8 & 7.90$^{\rm M13}$ & -2.50 & 8.10 & 0.6174\\
    WLM & 00 01 57.90 & -15 27 50.0 & 1.0 & 344.5 & 174.5$^{\rm O15}$ & 74$^{\rm O15}$ & 9.9 & 7.31$^{\rm LE19}$ & -2.13 & 8.15 & 0.1005 \\
    NGC~5068 & 13 18 54.80 & -21 02 20.7 & 5.2 & 217.4 & 342.4$^{\rm LA20}$ & 35.7$^{\rm LA20}$ & 6.0 & 9.36$^{\rm LE19}$ & -0.61 & 8.35 & 0.2729 \\
    NGC~7793 & 23 57 49.83 & -32 35 27.7 & 3.6 & 280.0 & 290$^{\rm LA20}$ & 50$^{\rm LA20}$ & 7.4 & 9.25$^{\rm LE19}$ & -0.65 & 8.35 & 0.0518\\
    NGC~1566 & 04 20 00.39 & -54 56 16.6 & 18 & 249.6 & 214.7$^{\rm LA20}$ & 29.6$^{\rm LA20}$ & 4.0 & 10.67$^{\rm LE19}$ & 0.60 & 8.57 & 0.0242 \\
    NGC~5236 & 13 37 00.95 & -29 51 55.5 & 4.9 & 386.5 & 225$^{\rm LA20}$ & 24$^{\rm LA20}$ & 5.0 & 10.41$^{\rm LE19}$ & 0.41 & 8.60 & 0.1770 \\
    \hline
    \end{tabular}
    \caption{Characteristics of sample galaxies (ordered by increasing gas-phase metallicity). The coordinates and redshifts are from the NASA/IPAC Extragalactic Database (NED). The optical radius $r_{\rm 25}$ is the radius of 25 mag arcsec$^{−2}$ at B band, taken from \citet{devaucouleurs1991third}. The T-Type is the indicator of morphological type, taken from HyperLeda \citep[][]{makarov2014hyperleda}. Distance is from the adopted distance in \citet{leroy2019z}.\\
    References for position angle (P.A.), inclination (Incl.), and \Mstar: M13=\citet{madden2013overview}, O15=\citet{oh2015high}, N17=\citet{namumba2017hi}, N18=\citet{namumba2018hi}, LE19=\citet{leroy2019z},  LA20=\citet{lang2020phangs} \\
    The position angle and the inclination are adopted from gas kinematics analysis (CO observation for large galaxies and \Hone\ observation for dwarf galaxies). \\
    $^{\dagger}$This work: SFR is derived from FUV+MIR \citep[][]{belfiore2023calibration} and \logOH\ is measured using Scal diagnostic \citep[][]{pilyugin2016new}. \\
    $^{\ast}$The MW extinction value is adopted from \citet{schlafly2011measuring} dust maps (see Sec.~\ref{sec:photometricdata}).}
    \label{tab:sample}
\end{table*}

Our sample consists of seven galaxies—three dwarf galaxies and four spiral galaxies—selected from the TYPHOON survey (see details in Sec.\ref{sec:typhoon}). These galaxies, listed in order of increasing gas-phase metallicity (Scal diagnostic; see Sec.~\ref{sec:scal}), are Sextans~A, NGC~6822, WLM, NGC~5068, NGC~7793, NGC~1566, and NGC~5236.

The selection criteria for this sample are the availability of the following observations: (1) IFS (TYPHOON survey-limited for wide field-of-view access), (2) CO (not strictly required for dwarf galaxies); (3) far-ultraviolet (FUV), near-infrared (NIR), mid-infrared (MIR) images; and (4) \Hone\ 21~cm data with a spectral resolution, better than 4\kms, to enable finding colder \Hone\ components. The characteristics of the sample galaxies are described in detail below and also in Tab.~\ref{tab:sample}.

\subsection{Data}
\label{sec:data}


\subsubsection{IFS - the TYPHOON survey}
\label{sec:typhoon}
The TYPHOON/PrISM survey is a pseudo-IFS galaxy survey for 44 nearby galaxies (D~$\lesssim$ 35~Mpc; Seibert et al. in prep.). This survey used the wide-field CCD imaging spectrograph of the du Pont 2.5~m telescope at Las Campanas Observatory in Chile. Using the step-and-stare technique (or Progressive Integral Step Method; PrISM) with a long-slit, the observations have a large FoV of approximately 18\arcmin $\times$\,(1.65\arcsec $\times \rm N$), where N represents the number of stepped slits. This enables a large coverage of nearly the entire optical disk of nearby galaxies.

The final data cubes, produced after data reduction, have the following key properties: (1) a wavelength range spanning 3650--8150~\AA, (2) a spectral resolution of $\Delta\lambda \sim$3.5\AA\ (corresponding to an FWHM of 8.24~\AA\ and R~$\sim$850 at 7000\AA), and (3) a spatial resolution of 1.65\arcsec, equivalent to a physical scale of 40~pc at a reference distance of 5~Mpc. Emission line fluxes in the spaxels are analysed using \texttt{lzifu} \citep[][]{ho2016lzifu}. Following the modelling and removal of the stellar continuum from each observed spectrum, \texttt{lzifu} fits a single, kinematically-aligned Gaussian to the emission lines and calculates their flux and error. For the intensities of [N~II]~6548 and [O~III]~4959, the \texttt{lzifu} assumes the intrinsic [N~II] and [O~III] intensity ratio fixed to the ratio given by quantum mechanics ([N~II]$\lambda$~6583 / [N~II]$\lambda$~6548 = 3.05 and [O~III]$\lambda$~5007 / [O~III]$\lambda$~4959 = 2.98) \citep[][]{storey2000theoretical}, as their ratios are independent of physical conditions. 


The observed emission line luminosity, used for further analysis, is corrected for dust extinction based on the Balmer decrement, $L_{\mathrm{H}\alpha, \rm obs}/L_{\mathrm{H}\beta, \rm obs}$. The corrected luminosity at wavelength $\lambda$, denoted as $L_{\lambda,\text{corr}}$ is expressed as:
\begin{equation}
    \label{eq:ldustcorr}
    L_{\lambda,\mathrm{corr}} = L_{\lambda}10^{0.4~k_{\lambda}~E(B-V)},
\end{equation}
and
\begin{equation}
    E(B-V) = \frac{\log \left( \frac{ L_{\text{H}\alpha,\text{obs}}/L_{\text{H}\beta,\text{obs}}}{L_{\text{H}\alpha,\text{int}}/L_{\text{H}\beta,\text{int}}} \right) }{0.4 \times (k (H\beta) - k (H\alpha))}, 
\end{equation}
where the intrinsic L$_{\text{H}\alpha,\text{int}}$/L$_{\text{H}\beta,\text{int}}$ = 2.86, under case B recombination with a temperature of T = 10,000 K and an electron density of $n_{\rm e}$ = 100 $\rm cm^{-3}$ \citep[][]{osterbrock1989astrophysics}. We assume $R(V) = 3.1$, based on the average MW extinction curve \citep{fitzpatrick1999correcting}. The extinction coefficient, $k(\lambda)$, is derived from the relation $k(\lambda) = $A($\lambda$)/E(B-V).

\subsubsection{CO data}
\label{sec:co}
We collect archival data cubes from CO low-J observations (either the J=2$\rightarrow$1 or J=1$\rightarrow$0 transition), depending on availability. For the spiral galaxies in our sample—NGC~5068, NGC~7793, NGC~1566, and NGC~5236—we use the CO (2-1) moment 0 maps (integrated intensity) and their associated uncertainty maps provided by the PHANGS-ALMA program \citep[][]{leroy2021phangs}. Of these four galaxies, all except NGC~7793 were observed with the ALMA array configuration of 12m+7m+TP (Total Power). The CO (2-1) transition for NGC~7793 was observed using the 7m array and TP.

For NGC~6822 and WLM, we use individual ALMA observations, obtained with the 7m-only configuration. For NGC~6822, the ALMA CO (1-0) observations (program codes: 2019.2.00110.S, PI: Kohno; 2021.1.00330.S, PI: Tosaki) consist of three separate mosaics covering the upper, middle, and lower regions of the galaxy’s main star-forming body. For WLM, we use ALMA CO (2-1) observations (program code: 2018.1.00337.S, PI: Rubio). The reduced, primary beam-corrected data cubes were downloaded from the ALMA data archive\footnote{\url{https://almascience.nrao.edu/aq/}}. We construct the integrated intensity map of CO (i.e., moment 0 map) using \texttt{CASA-immoments}, using the mask cubes, provided by the ALMA archive, which are created at the last stage of cleaning during the pipeline.

\subsubsection{Photometric data}
\label{sec:photometricdata}
We collect multi-wavelength photometric images from the DustPedia archive \citep[][]{clark2018dustpedia}, including GALEX FUV ($\lambda_{\rm eff} \sim 1528$~\AA), WISE W1 ($\lambda_{\rm eff} \sim 3.4\mu \rm m$), and WISE W4 ($\lambda_{\rm eff} \sim 22~\mu \rm m$), for five galaxies (Sextans~A, NGC~6822, NGC~7793, NGC~1566, and NGC~5236). The DustPedia archive provides Galactic reddening-corrected images for wavelengths shorter than 10$\mu$m, using the prescription of \citet{schlafly2011measuring}.
The galaxies used in this study are part of the Physics at High Angular resolution in Nearby GalaxieS (PHANGS) surveys. The PHANGS team provide public data products for the JWST and VLT/MUSE data on their website.

For the remaining two galaxies, WLM and NGC~5068, which are not included in the DustPedia project, we collect GALEX FUV band images taken as part of the GALEX Nearby Galaxy Survey \citep[][]{gildepaz2007galex} from the NED\footnote{DOI: 10.26132/NED1, URL: \url{https://ned.ipac.caltech.edu}} and WISE W1 and MIR W4 band images \citep[][]{wright2010wise} from the NASA/IPAC Infrared Science Archive (IRSA)\footnote{DOI: 10.26131/IRSA153, URL: \url{https://irsa.ipac.caltech.edu}}. Galactic extinction correction is applied to the FUV images using the E(B-V) (or A(V)) values from \citet{schlafly2011measuring}, in the same manner as for the DustPedia galaxies, obtained from the IRSA Galactic Dust Reddening and Extinction Service\footnote{DOI: 10.26131/IRSA537, URL: \url{https://irsa.ipac.caltech.edu/applications/DUST/}}.

The photometric images from DustPedia, NED, and IRSA are not background-subtracted. To correct for this, we construct a background image for each band separately by estimating the background level and small-scale fluctuations using the Python package \texttt{photutils.background}. We apply a 3$\sigma$ clipping to the image to determine the background level and its root-mean-square (RMS). The resulting background image is then subtracted from the original image.

We perform foreground star removal for the GALEX FUV and WISE W1 (3.4$\mu$m) images (and WISE W4 at 22$\mu$m if the image is substantially contaminated by foreground stars, visually inspected) by masking them out. We use the \texttt{PTS-7/8}\footnote{\url{https://github.com/SKIRT/PTS}} \citep[][]{verstocken2020high} software, a Python toolkit for the SKIRT radiative transfer code \citep[][]{camps2015skirt, camps2020skirt}. This software incorporates the 2MASS All-Sky Catalog of Point Sources \citep[][]{cutri20032mass} into the input image and identifies the peak intensity matched with the coordinates of each 2MASS source. The \texttt{PTS-7/8} then provides the mask with a size of the MW star-contaminating region around each point source, determined by (1) the coordinate matches and (2) the identification of the local peak within the small path of the FWHM of the PSF of each band image.

For NGC~6822, due to its close distance, the 2MASS source catalogue includes sources from both the galaxy and the Milky Way, necessitating an additional step to isolate and remove MW stars only. We convert the parallax information from the Gaia DR3 catalogue\footnote{DOI: 10.26131/IRSA541} \citep[][]{gaia2016gaia, gaia2023gaiadr3} into the distance and consider sources only within 200~kpc as foreground stars. The MW sources are then defined if the spatial separation with the coordinates from the Gaia DR3 source catalogue is less than 3\arcsec, chosen by the approximate standard deviation of the PSF for WISE W3 (6.5\arcsec) and WISE W4 (12\arcsec) bands. A mask map is then created for these MW stars, which is used for foreground star removal.

The uncertainty map for the photometric images is calculated based on a standard combination of Poisson noise, sky value scatter, and uncertainty in the mean sky. The uncertainty for a given wavelength band is given by the following expression \citep[see also][]{cook2014spitzer}:
\begin{equation}
    \sigma = \sqrt{f(\lambda) + A_{\rm ap} \times \sigma_{\rm sky}^2 + \frac{A_{\rm ap}^2 \times \sigma_{\rm sky}^2}{A_{\rm sky}}},
    \label{eq:uncertainty}
\end{equation}
where $f(\lambda)$ represents the number of photons per second detected (i.e., the Poisson error), $A_{\rm ap}$ is the area of the aperture size in pixel units squared (=resampled pixel size/native pixel size), $\sigma_{\rm sky}$ is the standard deviation of the sky values per native pixel, and $A_{\rm sky}$ is the area of the sky apertures in native pixel units squared. The values for $\sigma_{\rm sky}$ and $A_{\rm sky}$ are derived using the IDL tool, \texttt{sky.pro}.

\subsubsection{HI 21~cm data}
\label{sec:hi}
 
Part of our \Hone\ data are from archival Very Large Array (VLA) observations obtained from the THINGS \citep[for NGC~7793, NGC~5236; ][]{walter2008things} and LITTLE THINGS \citep[for Sextans~A, WLM; ][]{hunter2012little} surveys. For these four galaxies, the channel resolution is $\sim$~2.6~\kms. We use the `natural' weighted data, with beam sizes for the four sample galaxies ranging from 10.6\arcsec\ to 15.6\arcsec. For NGC~6822, we used Australia Telescope Compact Array (ATCA) data \citep[see ][]{deblok2000evidence}, with a channel resolution of 1.6~\kms. A recent deep data release from the MHONGOOSE survey \citep[][]{deblok2024mhongoose} using MeerKAT interferometer includes data cubes at various beam sizes (depending on different robust weighting schemes) for NGC~5068 and NGC~1566. We select data cubes with a beam size of approximately 14.1\arcsec\ $\times$ 9.7\arcsec\ and a channel resolution of 1.4~\kms\ \citep[see Table 4][]{deblok2024mhongoose}.

Tab.~\ref{tab:hiresolution} summarises the properties of the selected \Hone\ data, including the beam size, pixel size, physical resolution (converted from beam size at the distance of each galaxy), channel resolution, sensitivity in noise per channel, and their source paper.

\begin{table*}
    \centering
    \begin{tabular}{c|c|c|c|c|c}
    \hline
        Name & Beam (pixel) & Phys. res. & Chan. res. & Noise per chan. & Source \\
         & & pc & \kms\ & mJy/beam & \\
        \hline
        Sextans~A & 12\arcsec (3\arcsec) & 81 & 2.58 & 0.45 & H12 \\
        NGC~6822 & 43\arcsec (16\arcsec) & 104 & 1.6 & 3.9 & D06 \\
        WLM & 11\arcsec (3\arcsec) & 53 & 2.57 & 0.76 & H12 \\
        NGC~5068 & 14\arcsec (6\arcsec) & 353 & 1.38 & 0.171 & D24 \\
        NGC~7793 & 16\arcsec (4.5\arcsec) & 279 & 2.58 & 0.92 & W08 \\
        NGC~1566 & 13\arcsec(6\arcsec) & 1134 & 1.38 & 0.171 & D24 \\
        NGC~5236 & 16\arcsec (4.5\arcsec) & 380 & 2.57 & 0.83 & W08 \\
    \hline
    \end{tabular}
    \caption{The spatial/physical resolution, channel resolution of the \Hone\ data cube at which this study performs, and the sensitivity (noise per channel) of the original \Hone\ data cube. Every analysis except running the Gaussian decomposition tool is carried out with images resampled to beam size.\\
    The references for the \Hone\ data source: D06=\citet{deblok2006stellar}, W08=\citet{walter2008things}, H12=\citet{hunter2012little}, D24=\citet{deblok2024mhongoose}}
    \label{tab:hiresolution}
\end{table*}

\subsubsection{Convolution and resampling}
For the pixel-by-pixel analysis, we perform image convolution and resampling to match the spatial resolution of all datasets to the coarsest resolution, which corresponds to the \Hone\ beam size (Tab.~\ref{tab:hiresolution}) for all sample galaxies. We first apply image convolution to the GALEX FUV, WISE W1, WISE W4, and CO intensity maps (excluding \Hone\ data cubes). For GALEX FUV, and WISE bands, each has a unique point spread function (PSF), thus we use the kernel sets from \citet{aniano2011common}\footnote{\url{https://www.astro.princeton.edu/~draine/Kernels.html}} to transform the PSF of each instrument to a common Gaussian-shaped PSF matching the \Hone\ beam size. For CO intensity maps, we convolve them using the \texttt{CASA-imsmooth} assuming the Gaussian shape of the beam of original data cubes. Finally, we resample the smoothed data to the \Hone\ beam size using \texttt{SWARP}\footnote{\url{https://www.astromatic.net/software/swarp/}} \citep[][]{bertin2010swarp}.

The original \Hone\ data cubes are smoothed to have the symmetric Gaussian beam (Tab.~\ref{tab:hiresolution}), and resampled to have a pixel size of 1/3 to 1/4 of the beam size. The selection of the physical scale for each pixel is crucial, as larger pixels can smooth out the complex \Hone\ distribution in the velocity profile, hindering the detection of narrow \Hone\ components. After processing with the Gaussian decomposition tool, we resample the output maps (total, narrow, and broad \Hone\ components, see Sec.~\ref{sec:classification}) to match the resolution of other tracers for a consistent pixel-by-pixel analysis.


\section{Derivation of Physical Properties}
\label{sec:parameter}
\subsection{Molecular gas surface density ($\Sigma_{\rm mol}$)}
We convert the CO integrated intensity maps to H$_2$ surface density in \Msol pc$^{-2}$ using the following equations from \citet{leroy2021phangs}:
\begin{equation}
    \Sigma_{\rm mol} = \alpha_\mathrm{CO}^{1-0}~R_{21}^{-1}~I_\mathrm{CO (2-1)}~\mathrm{cos}~i,
\end{equation}
where $\alpha_{\rm CO}^{1-0}$ is the CO(1-0)-to-H$_2$ conversion factor, $R_{21}^{-1}$ is the flux ratio of CO(2-1) to CO(1-0), $I_{\rm CO (2-1)}$ is the integrated intensity of the CO (2-1) line in units of K\kms, and $i$ is the inclination of the galaxy. In this study, we assume a constant MW $\alpha_{\rm CO}^{1-0}$ of 4.35 \Msol pc$^{-2}$ (K\kms)$^{-1}$ for the entire galaxy sample. We adopt $R_{21}^{-1}$ = 0.65, following the assumptions by \citet{denbrok2021new} for CO (2-1) observations \citep[see also][]{leroy2021phangs}. For the CO (1-0) moment maps (used for NGC~6822), the term $R_{21}^{-1}~I_{\rm CO (2-1)}$ is replaced by $I_{\rm CO (1-0)}$.

\subsection{Star formation rate (SFR)}

We map out the SFR of galaxies at 100~Myr timescale, traced by the FUV band \citep[][]{hao2011dust, kennicutt2012star}. We combine it with a tracer of dust attenuation (MIR band, WISE W4 for this case), or the indicator of the amount of re-emitted emission from the absorption at a shorter wavelength by dust. The SFR from the two bands is:
\begin{equation}
    SFR_{\rm FUV+MIR}~[M_{\odot} {\rm yr}^{-1}] = C_{\rm FUV} L_{\rm FUV} + C_{\rm W4}^{\rm FUV}L_{\rm W4},
\end{equation}
where $C_{\rm FUV}$ and $C_{\rm W4}^{\rm FUV}$ are the conversion factors, and  $L_{\rm FUV}$ and $L_{\rm W4}$ are the luminosity at FUV and W4, respectively, in units of [erg s$^{-1}$]. The log$_{10}C_{\rm FUV}$ is -43.42 \citep[][]{leroy2019z}. For $C_{\rm W4}^{\rm FUV}$, we follow the \citet{belfiore2023calibration} method to take into account the possible overestimate of the standard hybrid SFR at the low SFR surface
density, low specific SFR (sSFR), low attenuation, and old stellar ages. Following the suggestions from \citet{belfiore2023calibration}, we include W1 (WISE 3.4$\mu$m) to correct the bias, depending on $Q = L_{\rm FUV}/L_{\rm W1}$, 
\begin{equation}
    {\rm log}~C^{FUV}_{W4} = 
    \begin{cases}
        a_0 + a_1 \times {\rm log_{10}}~Q ~~~~~~~ Q < Q_{\rm max}\\
        {\rm log_{10}}~C_{\rm max} ~~~~~~~~~~~~~~~~~Q > Q_{\rm max}
    \end{cases}
\end{equation}
Here, $a_0$ is ${\rm log_{10}}~C_{\rm max}-a_{1}~ {\rm log}~ Q_{\rm max}$, $a_1$ is 0.23$\pm$0.14, $Q_{\rm max}$ is 0.60$\pm$0.68, and log$_{10}C_{\rm max}$ is -42.73$\pm$0.12. We refer the readers to \citet{belfiore2023calibration} for details on the parameters. The integrated SFR for each galaxy is listed in Tab.~\ref{tab:sample}.

\subsection{Gas-phase metallicity}

\subsubsection{Star-forming H II region selection - BPT diagram}
\label{sec:bpt}
Different excitation sources affect the observed emission line ratios in nebular and galaxy spectra, enabling us to identify the dominant ionisation source-whether from star formation, Active Galactic Nuclei (AGN), shocks, or a combination of these (composite regions). To exclude composite- or AGN-dominated regions for more accurate gas-phase metallicity measurements from star-forming regions, we use the Baldwin-Phillips-Terlevich (BPT) diagram \citep{baldwin1981classification}. For a more conservative classification, we adopt the \citet{kauffmann2003host} line and classify pixels below this line as star-forming regions.

\subsubsection{Oxygen abundance}
\label{sec:scal}

In this study, we use the Scal metallicity diagnostic \citep[][]{pilyugin2016new}, an empirical method combined with a theoretical approach. The Scal calibration aligns with abundances derived from `direct' methods that use electron temperature ($T_{\rm e}$) measurements from (weak) auroral lines, with an accuracy of $\sim$0.1~dex across a broad metallicity range (7.0 $<$ \logOH $<$ 8.6). Additionally, it is not sensitive to gas pressure or ionisation parameters, making it particularly reliable for low-metallicity sources \citep[][]{devis2017using}.

The Scal calibration uses the following three ratios:
\begin{align}\label{eq:another}
\begin{split}
\mathrm{N}_2 &= (\mathrm{[NII]}\lambda 6548 + \lambda 6583) /\mathrm{H}\beta \,,\\
\mathrm{S}_2 &= (\mathrm{[SII]}\lambda 6716 + \lambda 6731) / \mathrm{H}\beta \,,\\
\mathrm{R}_3 &= (\mathrm{[OIII]}\lambda 4959 + \lambda 5007) / \mathrm{H}\beta \,.
\end{split}
\end{align}
and divides lower and higher branches defined by log N$_2$. For further details, we refer readers to \citet{pilyugin2016new}. The emission lines used in this study are corrected for dust extinction using Eq.~\ref{eq:ldustcorr}. The maps of the gas-phase metallicity derived here are presented in Sec.~\ref{sec:analysis}. We also derive the galaxy-integrated \logOH, by summing the intensities of relevant emission lines from the star-forming H~II regions in each galaxy, selected based on the BPT diagram (Sec.~\ref{sec:bpt}). The resulting galaxy-integrated \logOH\ values are listed in Tab.~\ref{tab:sample}.



\section{\Hone\ 21~cm line kinematic decomposition}
\label{sec:hi_decomposition}
\subsection{\baygaud}
\label{sec:baygaud}

\begin{figure*}
    \centering
    \includegraphics[width=0.94\linewidth]{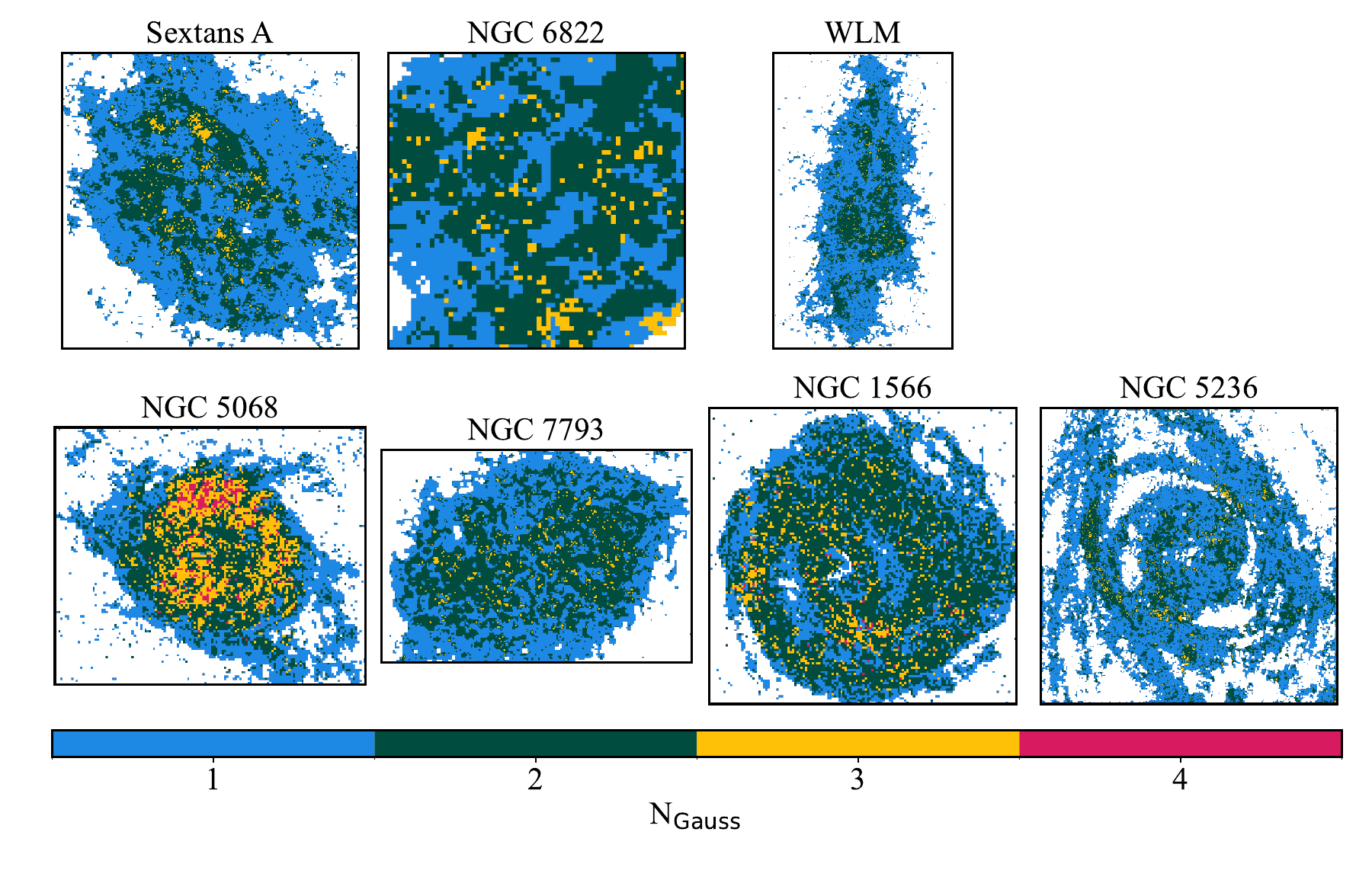}
    \caption{The map for the optimal number of \Hone\ Gaussian components (N$_{\rm Gauss}$) for our sample galaxies. NGC~5068 and NGC~1566 have up to four components due to the higher sensitivity and spectral resolution of the MHONGOOSE \Hone\ data that allow distinguishing fainter components (see Sec.~\ref{sec:baygaud}).}
    \label{fig:ngauss_map}
\end{figure*}

\baygaud\footnote{\url{https://github.com/seheon-oh/baygaud-PI}} \citep[][]{oh2019robust} is a robust kinematic decomposition tool for \Hone\ 21~cm data, employing the Bayesian Markov Chain Monte Carlo (MCMC) approach. For the line profiles of an input data cube (described in Sec.~\ref{sec:hi}), the model for the velocity profile can be expressed as:
\begin{equation}
    \label{eq:baygaud}
    G(x) = \sum_{i=1}^{N_{\rm max}} \frac{A_{i}}{\sqrt{2\pi}} \exp\bigg(\frac{-(x-\mu_{i})^{2}}{2\sigma_{i}^{2}}\bigg) + \sum_{j=0}^{n} b_{j}x^{j}\,,
\end{equation}
where G($x$) represents the sum of multiple Gaussian components, and $N_{\rm max}$ is the maximum number of Gaussians set by the user. We set $N_{\rm max} = 4$ for NGC~5068 and NGC~1566; $N_{\rm max} = 3$ for the other galaxies. The reason behind this (i.e., larger $N_{\rm max}$ for NGC~5068 and NGC~1566) is that their MHONGOOSE \Hone\ data cubes have higher spectral resolution and sensitivity relative to the other datasets to decompose the line profiles into more than three Gaussians. The parameters $A_{i}$, $\mu_{i}$, and $\sigma_{i}$ correspond to the amplitude, central velocity, and velocity dispersion of the {\it i}-th Gaussian component, respectively. $b_{j}$ is the coefficient of the $n^{\rm th}$ order polynomial used for the baseline fit. The tool then determines the optimal number of Gaussians for the profile by evaluating Bayes factors under the setting $N_{\rm max} =$ 3 or 4. For model selection of the optimal number of Gaussians, we adopt the `strong' model, defined as having a Bayes factor at least 10 times larger than the second-best model. For a detailed description of the tool, we refer the reader to \citet{oh2019robust}.

We apply two different signal-to-noise ratio (SNR) cuts for each Gaussian component: SNR$_{\rm amp}$ ($> 2$) and SNR$_{\rm area}$ ($> 2$). These criteria help avoid including spurious low-amplitude components that could be fitted, likely due to poorly defined background fitting, amplitudes lower than the noise level of the spectrum, and spiky components that have too narrow velocity dispersion to be counted as real signals. The SNR$_{\rm amp}$ is calculated as the amplitude of the Gaussian component divided by the noise level, which is measured from the $RMS (b)$ value of background-subtracted spectra (from baseline fitting in Eq.~\ref{eq:baygaud}). In the case of SNR$_{\rm area}$, the signal term is the integrated intensity, which is the total area under the Gaussian component, and the noise term is derived via:
\begin{equation}
    SNR_{\mathrm{area}, i} = I_{i} / (\sqrt{N_{\mathrm{chan}, i}} \times RMS (b)),
    \label{eq:integratedsn}
\end{equation}
where $I_{i}$ is the integrated intensity of a Gaussian component, $N_{\rm chan}$ is the number of channels associated with the Gaussian component. If any of the decomposed Gaussian components do not meet both criteria, we replace the multiple Gaussian fit results with the result from a single Gaussian fit \citep[][]{park2022gas}.

Fig.~\ref{fig:ngauss_map} presents the maps of the selected optimal number of \Hone\ Gaussian components (N$_{\rm Gauss}$) determined using a Bayesian approach that best describes the velocity profile. In general, the outskirts of galaxies are well-represented by a single Gaussian component, whereas star-forming regions and spiral arms show more complex velocity profiles that require multiple Gaussian components, consistent with previous studies \citep[e.g.,][]{braun1997temperature}.


\subsection{Classification of narrow/broad \Hone\ components}
\label{sec:classification}

\begin{figure*}
    \centering
    \includegraphics[width=0.94\linewidth]{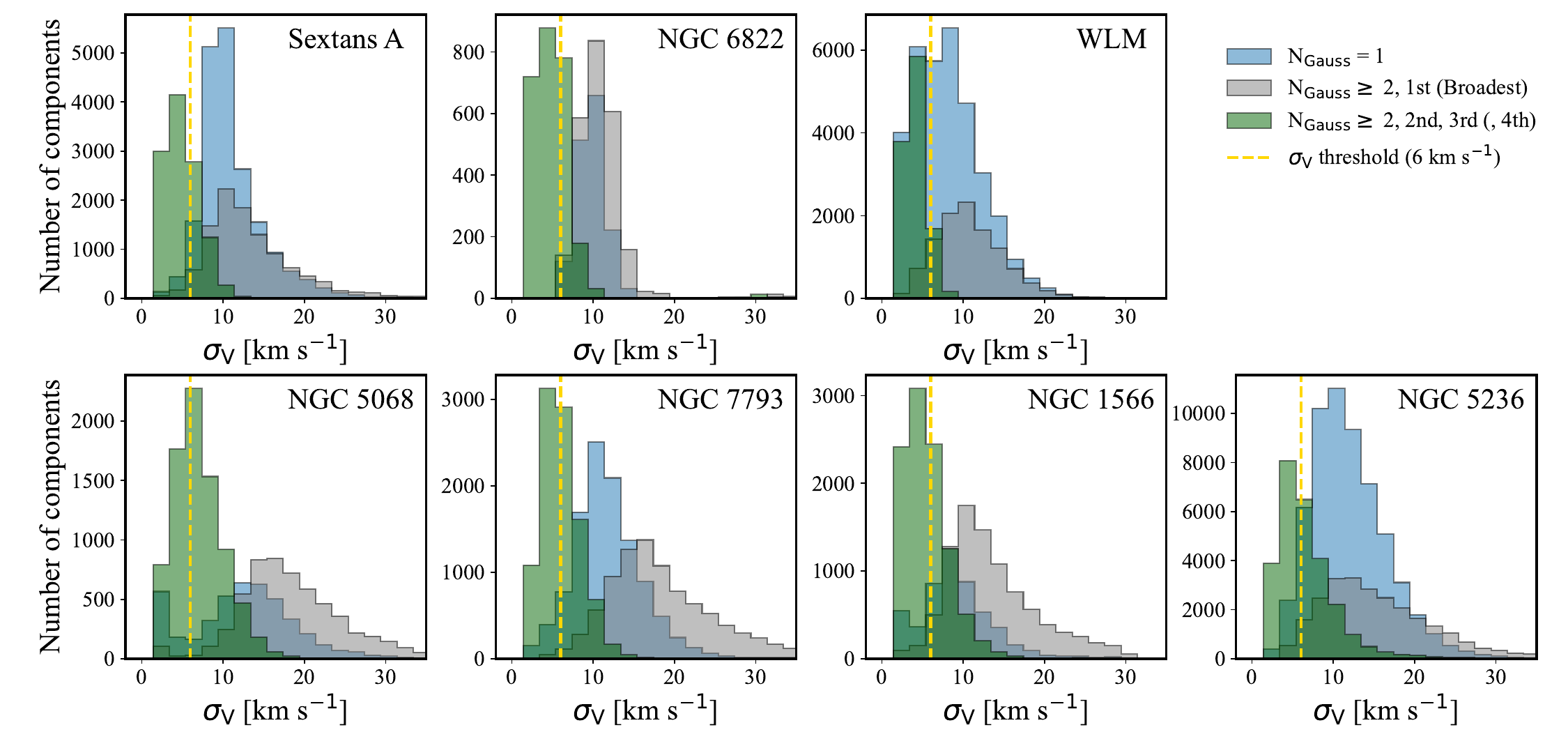}
    \caption{Histogram of velocity dispersion ($\sigma_{\rm V}$) for decomposed Gaussian components. The bin size is uniformly distributed from 1.4\kms (the lowest $\sigma_{\rm V}$ among the components across all sample galaxies) in steps of 2\kms\ up to 39.4\kms. The blue histogram represents the $\sigma_{\rm V}$ of single-Gaussian fitted components (i.e., N$_{\rm Gauss} = 1$). The grey and green histograms correspond to the broadest $\sigma_{\rm V}$ components and remaining (narrower) components from multi-Gaussian fitted profiles (i.e., N$_{\rm Gauss} \geq 2$). The yellow dashed line marks the velocity dispersion threshold (6\kms) used to classify broad and narrow components, which is applied to multi-Gaussian components.}
    \label{fig:veldisp_hist}
\end{figure*}

\begin{figure*}
    \centering
    \includegraphics[width=0.94\linewidth]{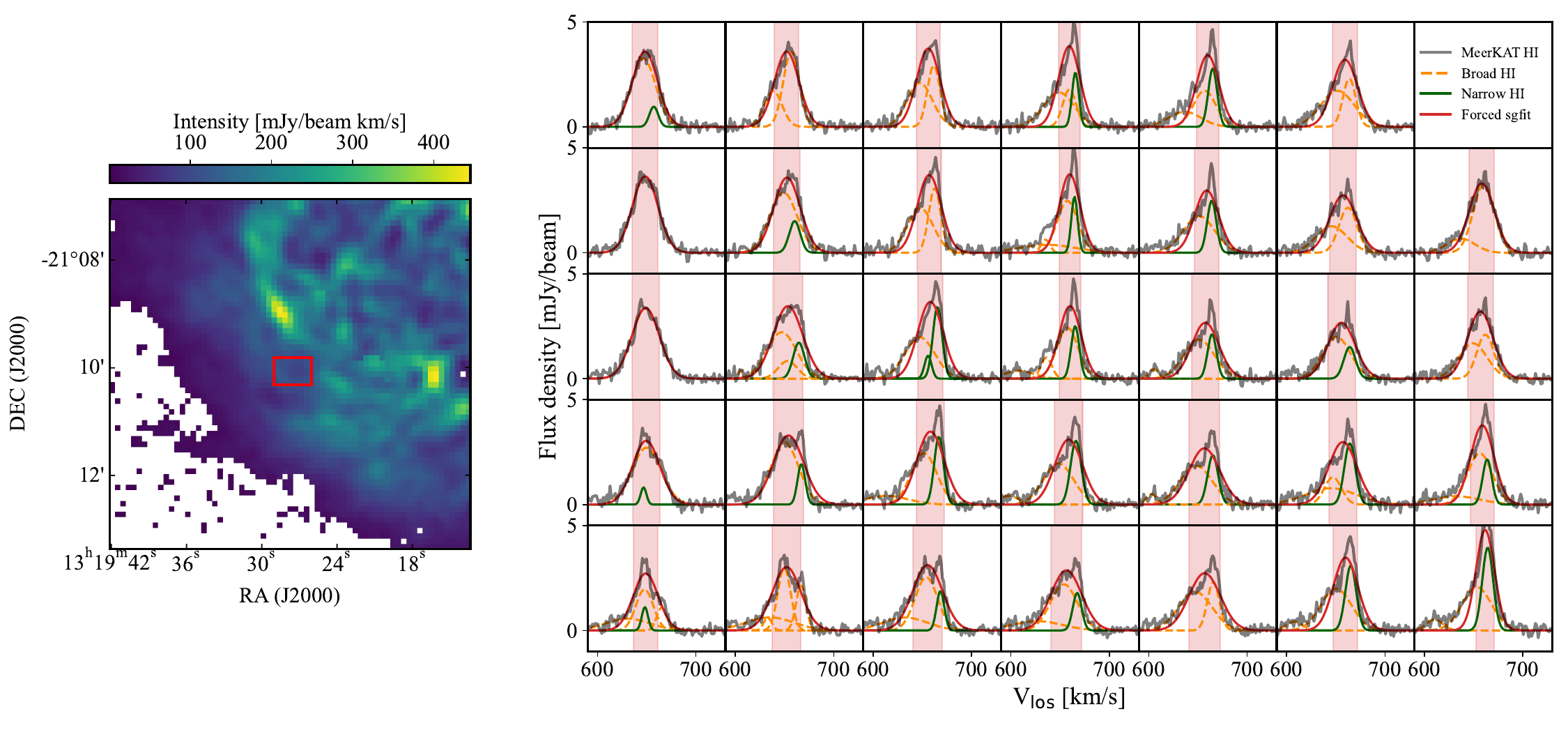}
    \caption{Demonstration of Gaussian fitting (\baygaud) results for the velocity profiles of an arbitrarily chosen region (red box on the left panel) of NGC~5068. The grey solid lines in the right panels show the velocity profiles of individual pixels. The red shaded area represents the V$_{\rm los}$ range within which colder, narrow \Hone\ components are identified, defined as $\mu_{\rm sg}$ $\pm$ $\sigma_{\rm sg}$ from the forced single Gaussian fit (red solid line). The orange dashed and green solid lines represent narrow and broad components, respectively, based on the criteria outlined in Sec.\ref{sec:classification}.}
    \label{fig:example_fit}
\end{figure*}

A velocity dispersion threshold is often used to classify cold \Hone\ components (see Sec.\ref{sec:introduction}). We take a cautious approach when defining our decomposed \Hone\ components as colder or warmer, being mindful that the CNM or cold \Hone\ has a maximum expected velocity dispersion of 2\kms, corresponding to a maximum kinetic temperature of T$_{\rm k} \sim$ 500~K. However, our ability to adopt 2\kms\ as the defining threshold for cold components is limited by the channel resolution of our \Hone\ data (1.4\kms\ to 2.6\kms). Thus, instead of directly identifying a cold \Hone\ component, we trace a phase of thermal condensation that transitions into cold \Hone, molecular gas or star formation, referring to it as the `colder' component. 

For velocity profiles best described by a single Gaussian component (i.e., N$_{\rm Gauss} = 1$), we exclude them from the colder \Hone\ component category, regardless of their velocity dispersion. In the case of multi-Gaussian decomposed profiles (i.e., N$_{\rm Gauss} \geq 2$), we exclude the broadest component from the colder \Hone\ category. These two criteria help to remove gas disk-induced profiles that do not show signatures of thermally condensing phases superimposed on the broad line. We define these excluded components as `broad'. For the remaining components, we apply a velocity dispersion threshold of 6~\kms, classifying those with velocity dispersions smaller than this value as colder components.

Fig.~\ref{fig:veldisp_hist} presents the velocity dispersion ($\sigma_{\rm V}$) histogram of all Gaussian components (after SNR cuts in the previous section) for the seven galaxies. The velocity dispersion values of the decomposed Gaussians range from the channel resolution of each data cube up to approximately 40~\kms. The broadest components (grey histogram) or single-Gaussian components (blue histogram) show a wide range of $\sigma_{\rm V}$ across most galaxies, whereas the narrower components (green histogram) are primarily constrained to $\sigma_{\rm V} < 10$\kms. We adopt a velocity dispersion threshold of 6\kms, indicated by the yellow dashed line.

Additionally, we classify components with $\sigma_{\rm V} <$ 6~\kms as colder \Hone, only if they align with the bulk motion of the galaxy, ensuring that components unrelated to star formation and those outside the galactic disc are excluded. To achieve this, we disregard Gaussian components whose central velocity ($\mu_{i}$ in Eq.~\ref{eq:baygaud}) deviates significantly from the bulk motion of the \Hone\ disc. The bulk motion is defined using the central velocity ($\mu_{\rm sg}$) and velocity dispersion ($\sigma_{\rm sg}$) from a forced single-Gaussian fit (sg), which models all velocity profiles with a single Gaussian component only\footnote{The single Gaussian fitting result is obtained simultaneously with the optimal Gaussian fitting using \baygaud.}. Specifically, if a narrow component from a multi-Gaussian fit has a central velocity ($\mu_{i}$) outside the range $\mu_{\rm sg} \pm \sigma_{\rm sg}$, we exclude it from the colder \Hone\ category and instead classify it as broad.

Fig.~\ref{fig:example_fit} presents an example of the optimal Gaussian fitting results for an arbitrarily chosen region in NGC~5068. The figure demonstrates that \baygaud\ effectively models the observed velocity profile using either multiple or single Gaussian components. The final criterion for classifying colder \Hone\ components—alignment with the bulk motion of the \Hone\ gas—is also visualised. This is evident from several narrow components located outside the red-shaded area, which represents the range $\mu_{\rm sg} \pm \sigma_{\rm sg}$. We find some discontinuities in the Gaussian decomposition results between adjacent pixels, occasionally missing colder \Hone\ components due to the definition criteria (i.e., the velocity dispersion threshold, $\sigma_{\rm V} <$ 6~\kms), even when neighbouring pixels contain colder \Hone\ gas. This issue can be mitigated by incorporating spatial coherence when decomposing velocity profiles into multiple Gaussians, as implemented in tools like \texttt{ROHSA} \citep[][]{marchal2019rohsa}. However, such an approach is beyond the scope of this study. We acknowledge this as a caveat of our analysis.

In summary, our colder \Hone\ components are defined by the following step-by-step criteria:
\begin{itemize} 
    \item[1.] The velocity profile is decomposed into two or more Gaussian components (i.e., N$_{\rm Gauss} \geq 2$).
    \item[2.] The distinct Gaussian component is not the broadest in the same velocity profile.
    \item[3.] The distinct Gaussian component has a velocity dispersion smaller than 6~\kms.
    \item[4.] The velocity of the component aligns with the bulk motion of the \Hone\ gas of each velocity profile, as defined by the forced single Gaussian fit of the profile.
\end{itemize}
For simplicity, we interchangeably use the terms `colder' and `narrow' \Hone\ components. All other components (i.e., broadest or best fit by a single Gaussian or components with $\sigma_{\rm V} > 6$\kms) are classified as `broad' \Hone\ components.

\section{Relations between Narrow \Hone, Molecular Gas, and SFR}
\label{sec:analysis}
\subsection{Narrow and Broad \Hone\ Components: Their Link to Molecular Gas and Star Formation}



\begin{figure*}
    \centering
    \includegraphics[width=0.98\linewidth]{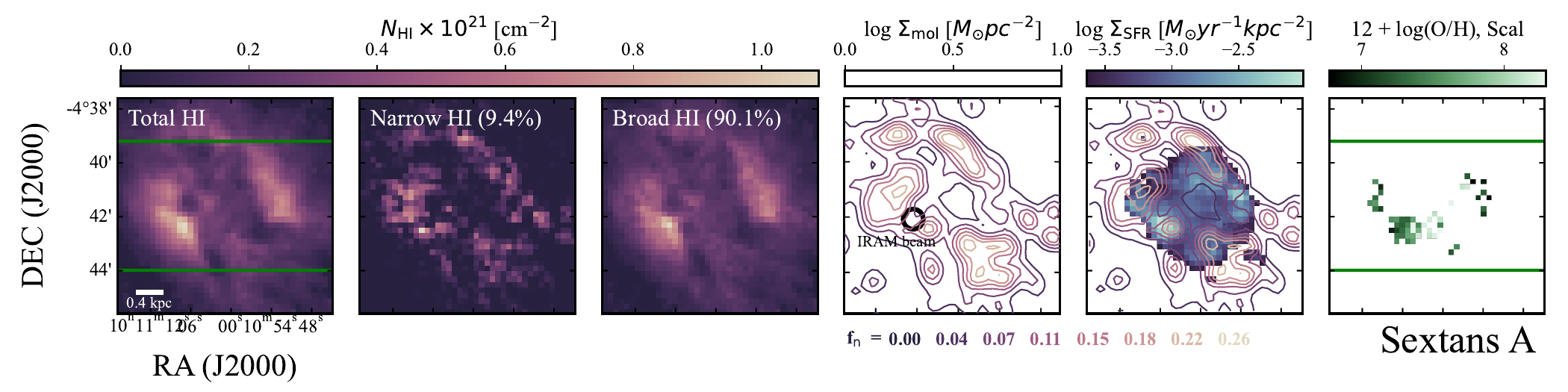}
    \includegraphics[width=0.98\linewidth]{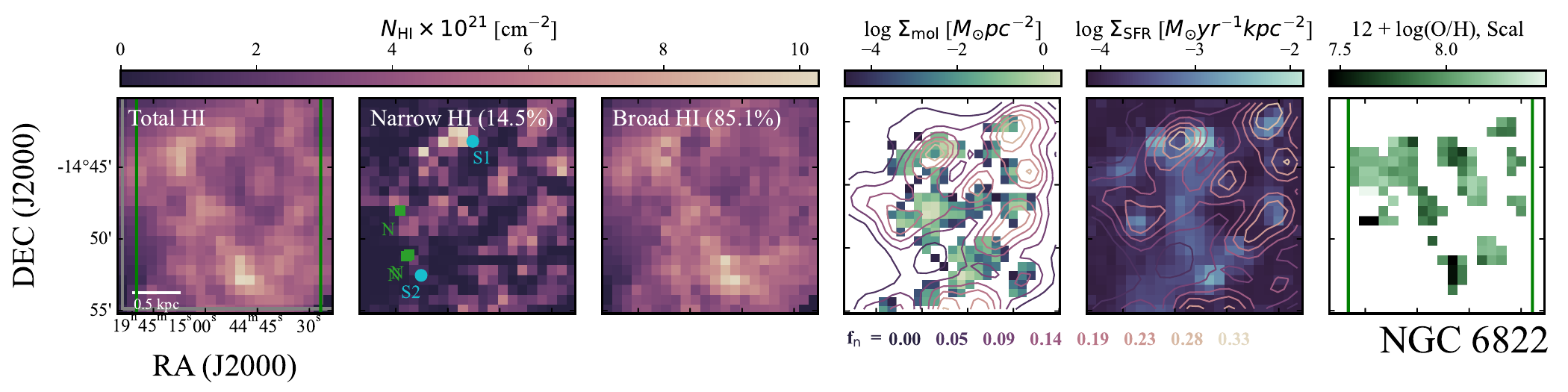}
    \includegraphics[width=0.98\linewidth]{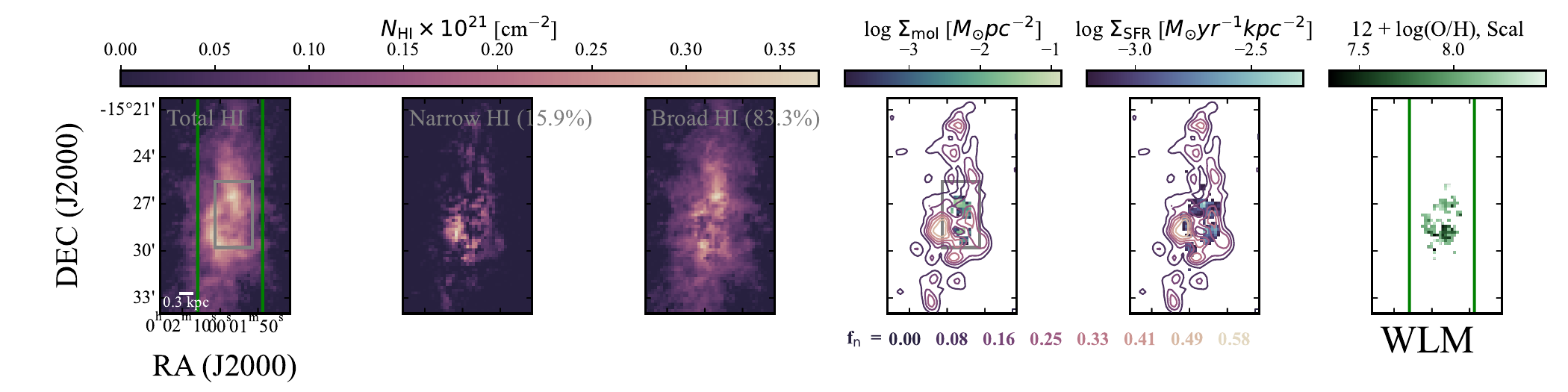}
    \caption{Left three panels: \Hone\ column density maps of total, narrow and broad \Hone\ components of dwarf galaxies, assuming optically thin medium. The mass fraction relative to the total \Hone\ is shown next to the phase name.  Fourth and fifth panels: Molecular gas surface density map (where available) in units of \Msol pc$^{-2}$ and SFR surface density map in units of \Msolyr kpc$^{-2}$, with the contour overlaid at \fn\ values indicated below the panels. Rightmost panel: Gas-phase metallicity map derived using Scal diagnostics. The grey boxes and green lines in the total \Hone\ and metallicity map indicate the approximate FoV of CO and TYPHOON IFS observations, respectively. The galaxy name is shown on the bottom right of each panel. For NGC~6822, the background sources with \Hone\ absorption detection and non-detection are indicated as S1 or S2 and N, respectively. The north is up and the east is left.}
    \label{fig:maps_dwarf}
\end{figure*}

\begin{figure*}
    \centering
    \includegraphics[width=0.98\linewidth]{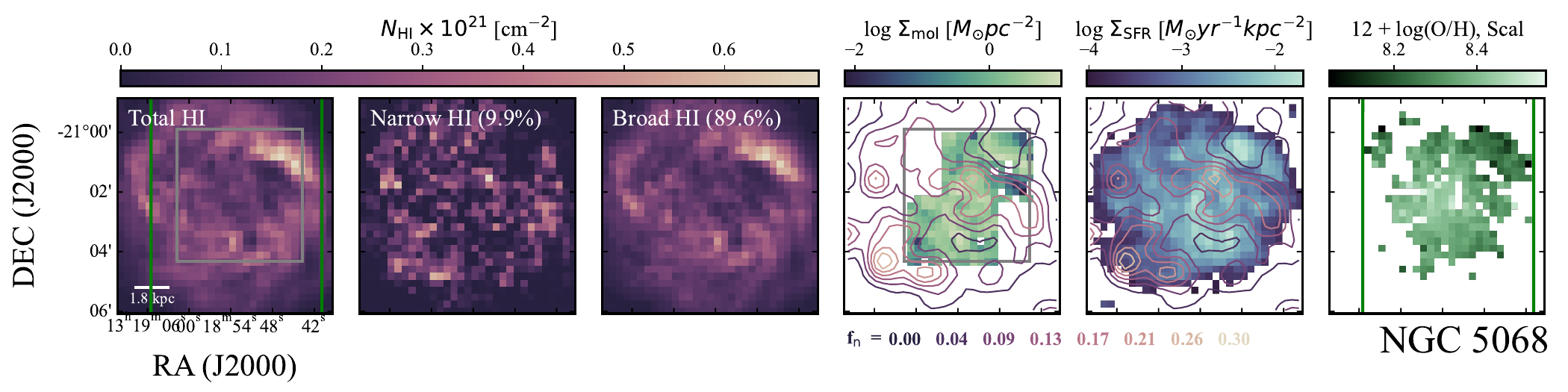}
    \includegraphics[width=0.98\linewidth]{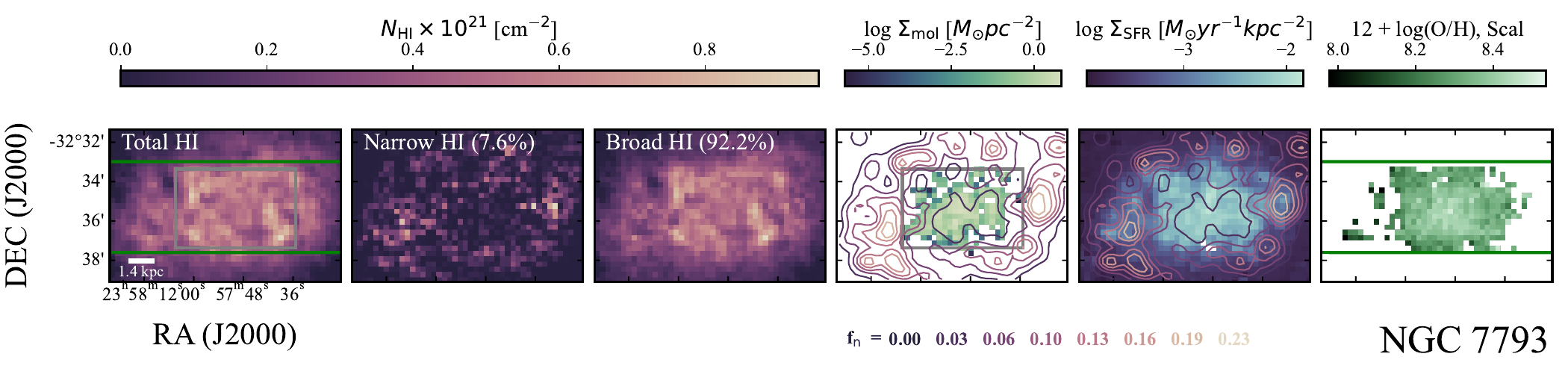}
    \includegraphics[width=0.98\linewidth]{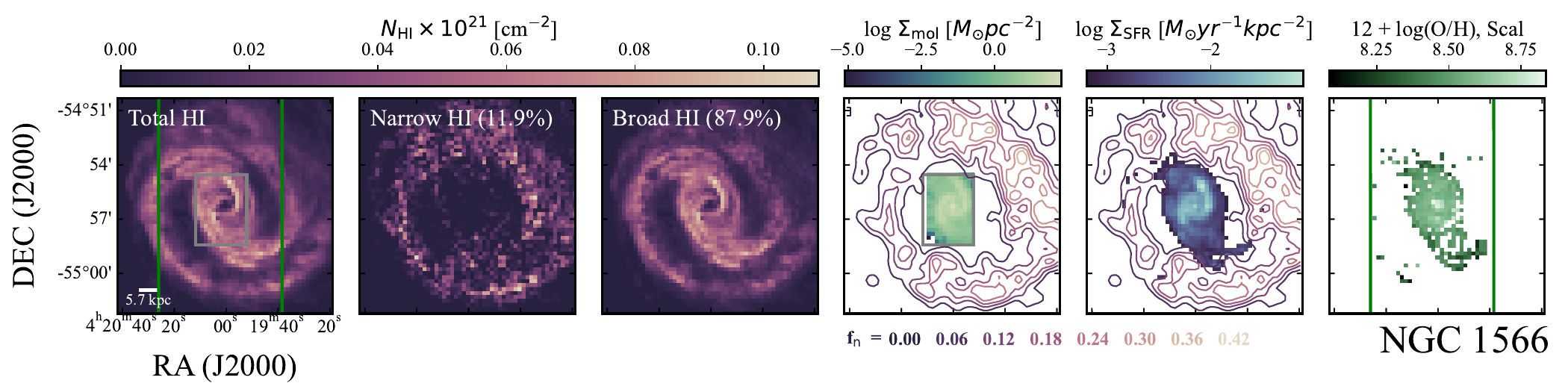}
    \includegraphics[width=0.98\linewidth]{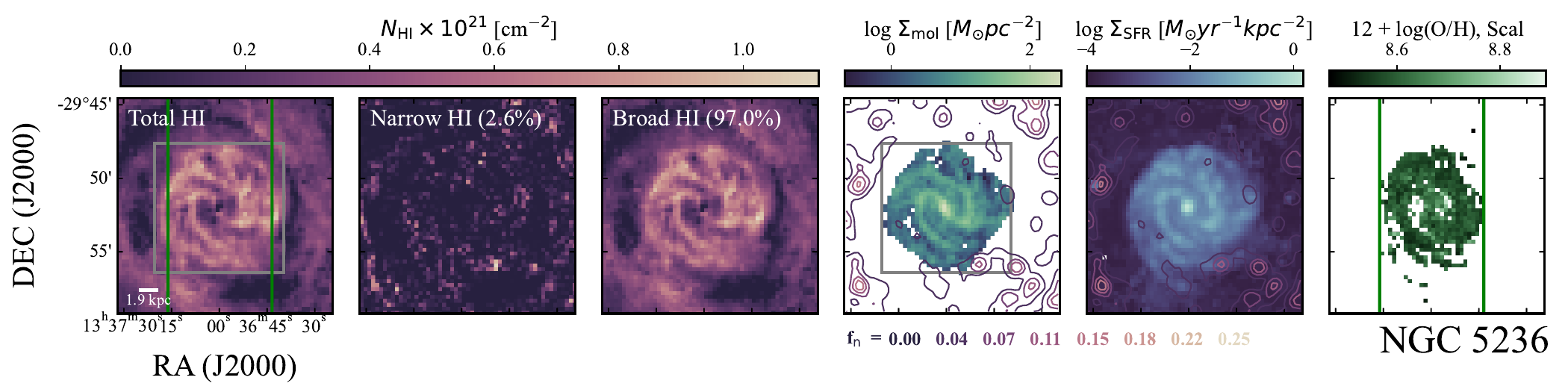}
    \caption{Same as Fig.~\ref{fig:maps_dwarf}, but for spiral galaxies.}
    \label{fig:maps_spirals}
\end{figure*}

The maps of \Hone\ column density for the total, narrow, and broad \Hone\ components are shown in the left three panels of Fig.~\ref{fig:maps_dwarf} and Fig.~\ref{fig:maps_spirals}, for three dwarf galaxies and four spiral galaxies, respectively. The \Hone\ column density is derived from the flux density ($S_{\nu}$) in Jy~beam$^{-1}$\kms, assuming optically thin conditions for \Hone, as outlined below:
\begin{equation}
    N_{\rm HI}~[\mathrm{cm}^{-2}] = 1.83~\times~10^{18} \int T_{B}~dV,
\end{equation}
where 
\begin{equation}
    T_{\rm B}~[\mathrm{K}] = 1.222~\times~10^{6} \frac{S_{\nu}}{\nu^{2}b_{\rm maj}b_{\rm min}}.
\end{equation}
Here, $\nu$ represents the rest frequency in GHz ($\nu = 1.4204$ GHz for the 21~cm hyperfine transition), while $b_{\rm maj}$ and $b_{\rm min}$ correspond to the beam sizes along the major and minor axes, respectively, in arcseconds.

We find that colder \Hone\ components make only minor contributions to the total \Hone\ content, comprising 9\%, 14\%, 16\%, 10\%, 8\%, 11\%, and 3\% for Sextans~A, NGC~6822, WLM, NGC~5068, NGC~7793, NGC~1566, and NGC~5236, respectively. Based on the column density maps of the total, narrow, and broad \Hone\ components in Fig.~\ref{fig:maps_dwarf} and Fig.~\ref{fig:maps_spirals}, we find that, in general, the narrow and broad \Hone\ components have distinct spatial distributions across galaxies. Broad components are spread throughout the entire gas disc, showing a distribution similar to that of the total \Hone\ for two main reasons: 1) broad components are naturally the primary contributors to the velocity profile, as per our classification method, and 2) profiles in the outer regions of the galaxy, where the \Hone\ signal is less complex, are decomposed into $N_{\rm Gauss} = 1$, which is integrated into the broad category. On the other hand, general features of narrow \Hone\ components are the clumpy and filamentary structures in our dwarf and spiral galaxies.

In the fourth and fifth panels of Fig.~\ref{fig:maps_dwarf} and Fig.~\ref{fig:maps_spirals}, we present the surface density maps of molecular gas and SFR, $\Sigma_{\rm mol}$ (traced by CO) and $\Sigma_{\rm SFR}$ (from FUV+MIR), respectively, overlaid with contours of the narrow gas fraction (\fn):
\begin{equation} 
    f_{n} = I_{\rm narrow HI}/I_{\rm total HI}\,.
\end{equation}
where $I_{\rm narrow HI}$ and $I_{\rm total HI}$ are the intensity of narrow \Hone\ and total \Hone. Below, we describe the spatial distribution of the total, narrow, and broad \Hone\ components, as well as the morphological relationship between \fn\ (contours), molecular gas content, and SFR in each galaxy.

\subsubsection{Dwarf galaxies - Sextans~A, NGC~6822, and WLM}

Sextans~A contains an \Hone-deficient centre, characterised by a large hole ($\sim$1.5~kpc in diameter) and concentrated \Hone\ distributions along the southeast (SE) and northwest (NW) edges \citep[][]{skillman1988hi}. One possible explanation for this feature is that stellar feedback has displaced neutral gas toward the periphery, depleting the central region \citep[][]{skillman1988hi, vandyk1998recent}. The narrow \Hone\ components in Sextans~A appear clumpier than the broad (or total) \Hone, highlighting high-density regions in the southwest (SW), northeast (NE), and along the SE and NW edges. This clumpiness is particularly evident in the \fn\ contours shown in the fourth and fifth panels.

A single CO detection has been reported in this galaxy using the IRAM-30m single dish \citep[][]{shi2015weak}, marked by a black circle in the fourth panel. This region also shows a high density of recent star formation (on the rightmost panel). The CO-detected area appears deficient in narrow \Hone, possibly due to molecular gas formation depleting the \Hone\ reservoir through thermal condensation. Alternatively, ongoing star formation may ionise the gas and/or broaden the 21~cm line through stellar feedback. The four high-\fn\ clumpy regions (each on SE, NW, SW, and NE edges of the galaxy) coincide with active star-forming regions identified by \citet{vandyk1998recent}, \citet{dohmpalmer2002deep}, and \citet{garcia2019ongoing}, suggesting that colder \Hone\ gas may contribute to star formation.

NGC~6822 features a stellar bar extending from the northeast to the southwest, accompanied by several distinct star-forming regions. In the total \Hone\ map, high \Hone\ concentrations generally align with CO detections and recent star formation, even before applying the \Hone\ kinematical decomposition. However, in some star-forming regions, the classified narrow \Hone\ clumps exhibit a stronger morphological correlation. For instance, in two star-forming regions on the eastern and western sides of the galaxy, \fn\ follows a clumpy distribution, peaking at locations with high $\Sigma_{\rm SFR}$.

In the narrow \Hone\ component map, we show the two absorption line detections by \citet{pingel2024local}, indicated as S1 (NVSS J194452-144311) and S2 (NVSS J194507-145233) and three non-detections. They measured the spin temperature of the two absorption lines and the CNM fraction over total \Hone, deriving $f_{\rm CNM}$ = 0.330 and 0.115 for S1 and S2, respectively. We do not see a spatial correlation between the absorption line detections and the high-density narrow \Hone\ gas regions. S1 is located at high molecular gas density and a high SFR region, supporting the existence of cold gas in the neutral phase in the star-forming regions. The high narrow \Hone\ density (or high \fn) region on the north side of the galaxy is close to S1, spatially offset by $\sim$~100~pc. S2 does not coincide with the narrow \Hone\ component, nor with the molecular gas and SFR distributions. We also include non-detections (indicated as `N'), exhibiting no particular spatial correspondence with narrow or broad components. Given the limited background sources for absorption line detections, a fair analysis of CNM and narrow \Hone\ components is challenging, preferring in closer systems, such as MW and Magellanic Clouds \citep[e.g., ][]{dempsey2022gaskap, nguyen2024local, lynn2025considerations}.

In WLM, narrow components form a distinct clump on the eastern side of the galaxy, spanning $\sim$400~pc. CO clouds are detected and identified by \citet{archer2022environments} in star-forming regions and mapped in the fourth panel of Fig.~\ref{fig:maps_dwarf}. No particular spatial correlation between \fn\ and molecular gas density is found, with several regions with low \fn\ containing high molecular gas density. Given the galaxy's high inclination ($\sim$~74\degree), we note that the velocity profiles classified as broad may include multiple narrow \Hone\ components along the line of sight. The inclination effects are discussed in Sec.~\ref{sec:discussion-limitations}.

\subsubsection{Spiral galaxies - NGC~5068, NGC~7793, NGC~1566, and NGC~5236}
In our sample of spiral galaxies, the distribution of narrow \Hone\ gas differs noticeably from that of molecular gas and SFR. This is in contrast to what we observe in our dwarf galaxy sample, where the distributions of narrow \Hone\ components, molecular gas, and SFR are spatially more closely related. This discrepancy suggests that the processes driving the star formation are more closely linked with molecular rather than colder \Hone\ compared to dwarf galaxies.

NGC~5068 shows prominent arc-like features, particularly bright in \Hone\ (both total and broad) on the northwestern side of the galaxy. These features align with the galaxy's spiral arms, where both molecular gas and SFR surface densities are high. In contrast, the narrow \Hone\ component is more dispersed across the galaxy, with multiple peaks extending toward larger galactocentric radii. Due to the limited FoV of the CO observations for NGC~5068 (indicated by the grey box in the leftmost panel), establishing a clear morphological connection between \fn\ and the molecular gas distribution is challenging. In the central regions of the galaxy, where \fn\ is high, the narrow \Hone\ gas does not appear to be correlated with either the molecular gas distribution or SFR.

The flocculent spiral galaxy NGC~7793 shows a similar scattered distribution of narrow \Hone\ components extending to larger galactocentric radii, as seen in NGC~5068. While the spiral structures are not prominent in the narrow \Hone\ distribution, its high-density regions are observed in the outskirts of the galaxy (R$>$ R$_{25}$), with two noticeable features on the eastern and western sides of the galaxy. Due to the deficiency of narrow \Hone\ in the centre region of NGC~7793, where CO observations are available, and CO observations lacking in the outer regions of the galaxy, it is challenging to determine its relationship with molecular gas density. When comparing \fn\ with the SFR surface density at large galactocentric radii, some high \fn\ peaks coincide with regions of elevated SFR relative to their surroundings.

The high-metallicity spiral galaxies NGC~1566 and NGC~5236 feature grand-design spiral arms visible in all gas phases-total, narrow, broad, and molecular gas-and in SFR maps. Both galaxies show \Hone\ deficiency in their centres (R $<$ 2~kpc), which could be caused by: 1) the transition of \Hone\ to molecular gas and star formation, or 2) the influence of the AGN in the centre, which may ionise the neutral gas or push it outward to surrounding regions.

In NGC~1566, narrow \Hone\ components are notably deficient in the central region (R$ < 10$~kpc) but become more abundant at the galaxy's outskirts along the spiral arms. The asymmetry in \Hone, particularly the outskirts of the galaxy, is observed in the total \Hone\ map \citep[see also][]{elagali2019wallaby, maccagni2024mhongoose}. Specifically, the \Hone\ on the western side of the galaxy shows a smoothly decreasing density towards higher radii, while on the eastern side, the density drops sharply at the edge. \citet{maccagni2024mhongoose} compiled multiple \Hone\ data cubes with different spatial resolutions and sensitivities, showing that the asymmetric \Hone\ gas disc is possibly the result of 1) ram pressure stripping \citep[][]{elagali2019wallaby} or 2) past interaction between NGC~1566 and NGC~1581, which stripped diffuse gas from the outskirts of the galaxy's eastern side. The distribution of \fn\ shows a significantly asymmetric morphology at larger galactocentric distances (R$ > 15$~kpc) as well as the total \Hone. Notably, the \fn\ is much lower on the northwestern edge compared to the southeastern edge. If ram pressure stripping was at play, this high \fn\ could support the existence of multi-phase ram pressure-stripped gas \citep[e.g.,][]{choi2022ram}, along with molecular gas present in the stripped tail \citep[e.g.,][]{lee2018alma, brown2021vertico}. Or, if the interaction with NGC~1581 in the past is the culprit that reshaped the diffuse \Hone\ distribution in the outskirts of the galaxy, the gas may be prevented from being condensed towards colder \Hone\ and molecular gas. Since CO observations are only available for the inner region, a direct spatial comparison between narrow \Hone\ and molecular gas content is not feasible. 


In NGC~5236, the narrow \Hone\ gas appears clumpy and sometimes filamentary along the spiral arms. However, due to the central \Hone\ deficiency, performing a morphological analysis—especially in comparison with molecular gas surface densities—becomes challenging.

\subsection{Correlation at different spatial scale}
\label{sec:cross-correlation}
\begin{figure*}
    \centering
    \includegraphics[width=0.32\linewidth]{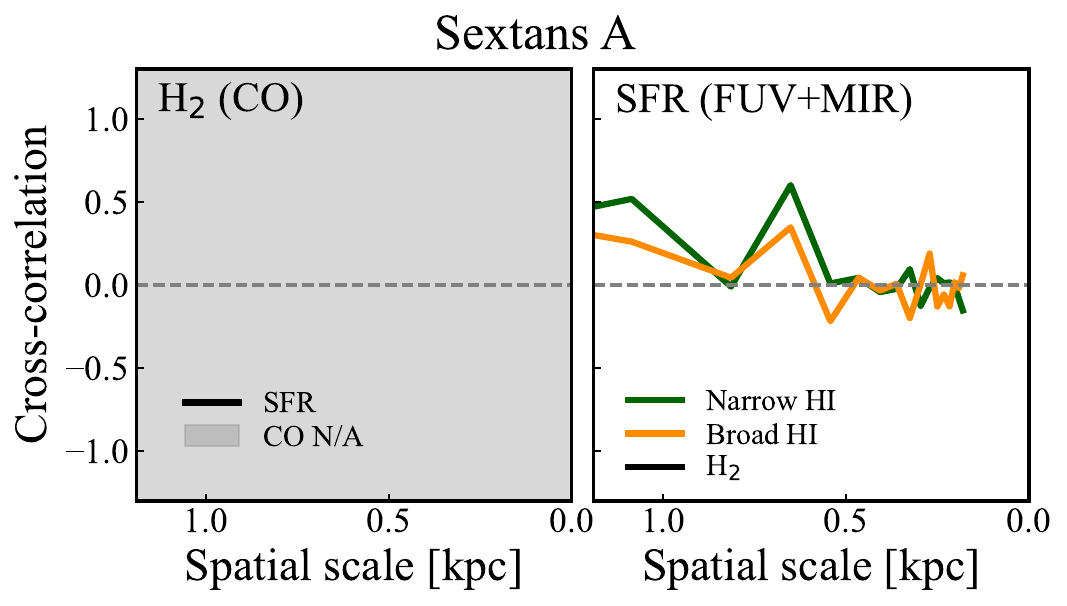}
    \includegraphics[width=0.32\linewidth]{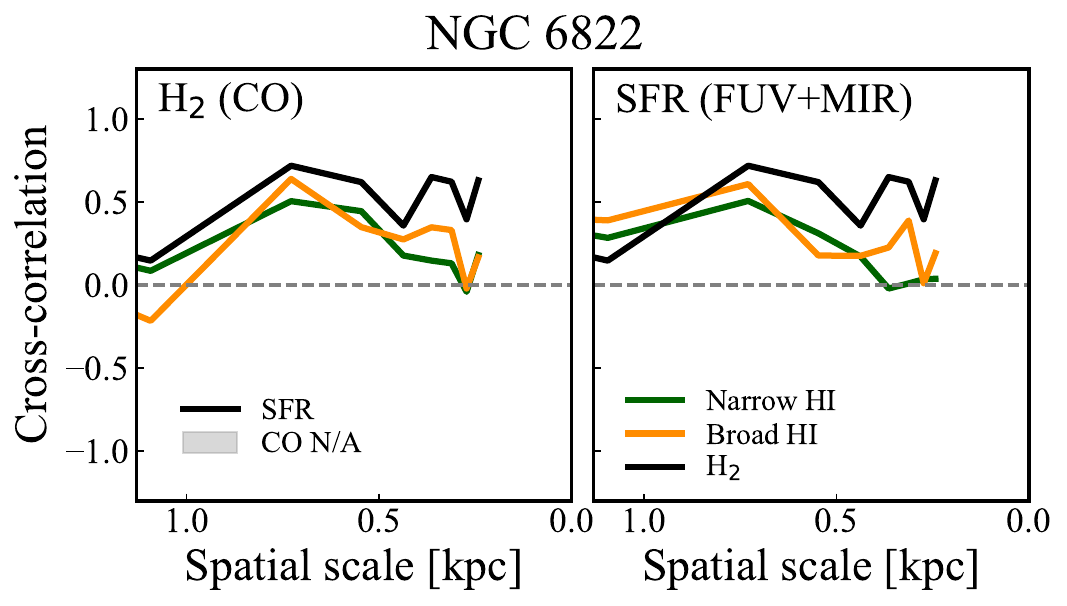}
    \includegraphics[width=0.32\linewidth]{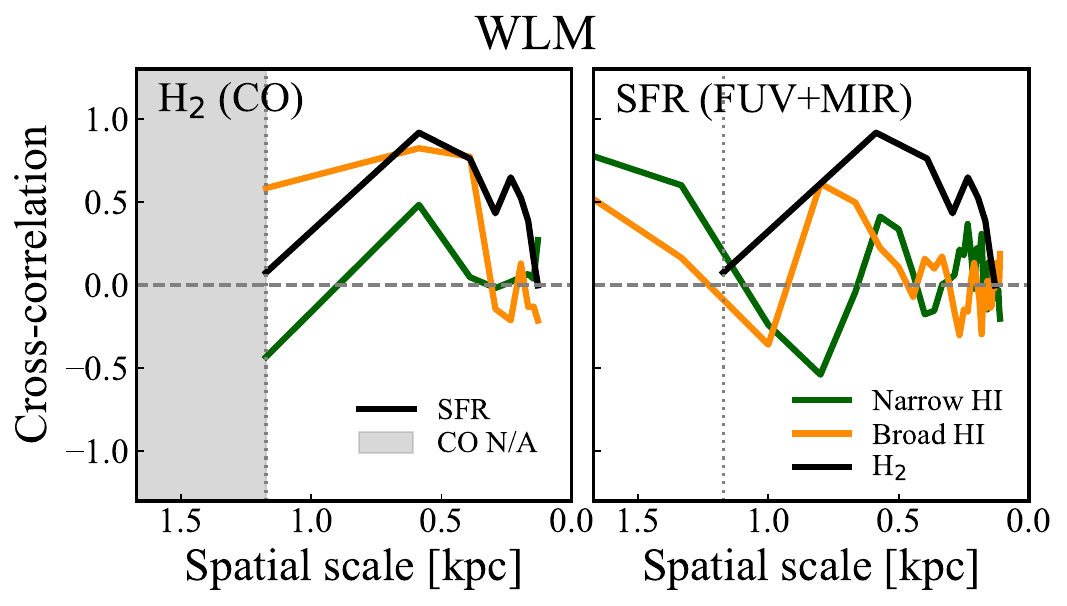}
    \includegraphics[width=0.32\linewidth]{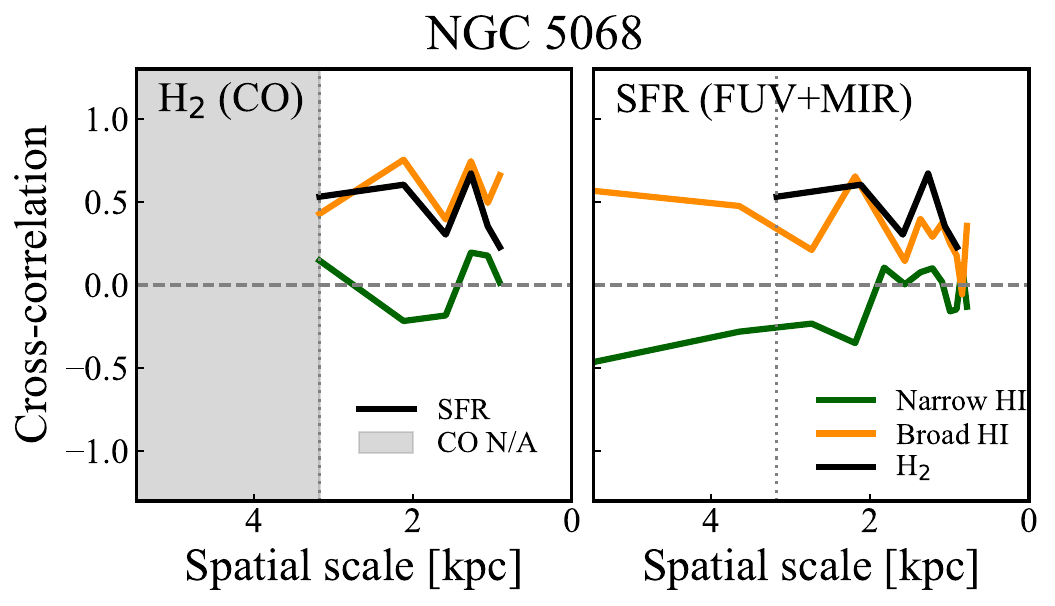}
    \includegraphics[width=0.32\linewidth]{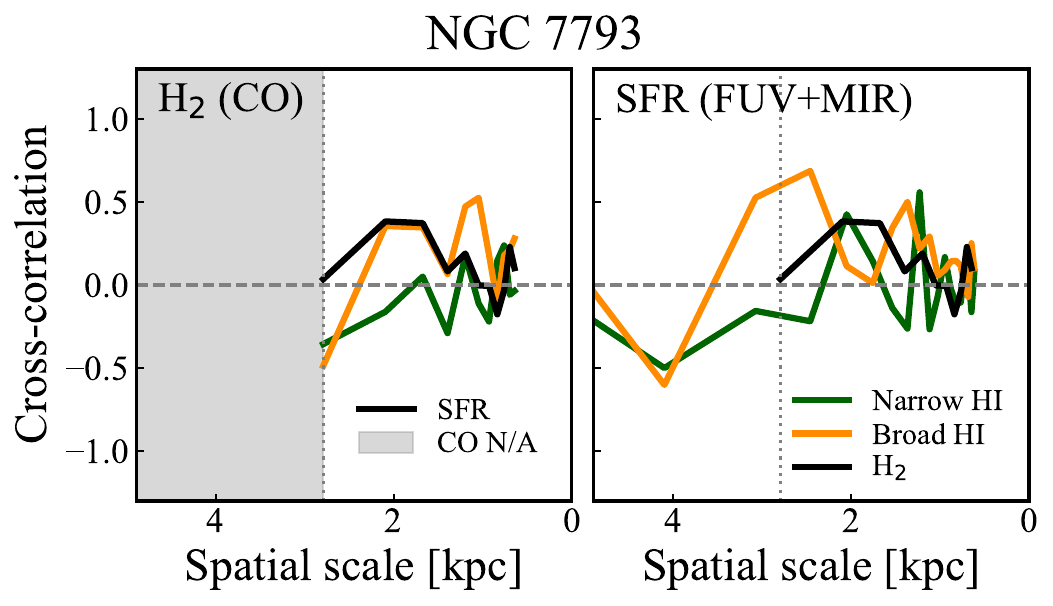}
    \includegraphics[width=0.32\linewidth]{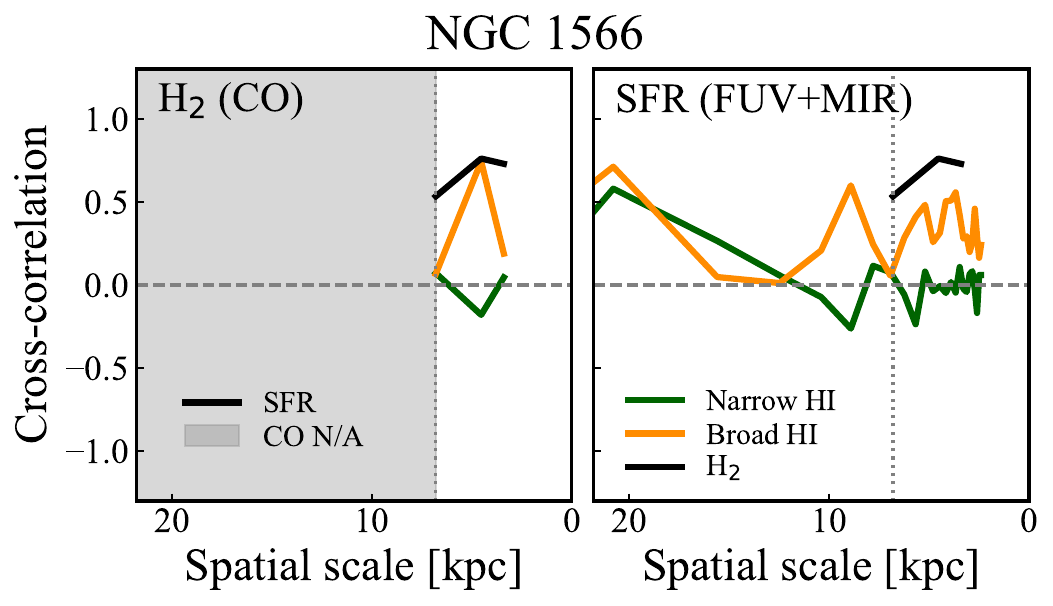}
    \includegraphics[width=0.32\linewidth]{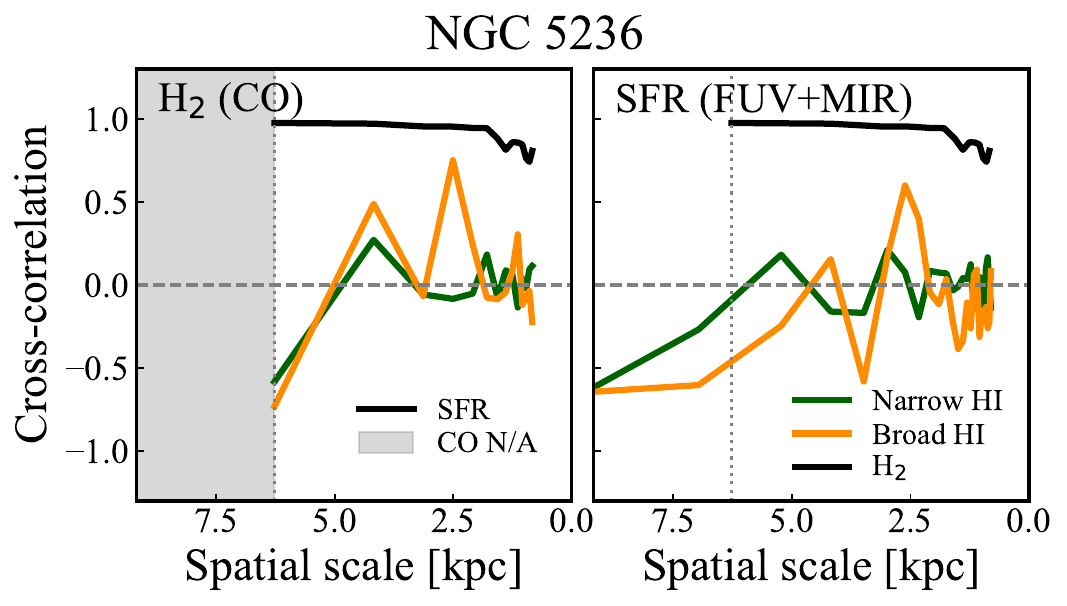}
    \caption{The cross-correlation analysis for the narrow (green) and broad (orange) \Hone\ components with the molecular gas surface density (left panels) and SFR surface density (right panels) of our sample galaxies. The black solid lines show the cross-correlation between molecular gas content and SFR in both panels. The x-axis represents the spatial scale, $k$, in units of kpc, ranging from the optical radius ($r_{25}$; left side) of each galaxy to the pixel size (right side). The x-ranges differ between galaxies due to the varying physical resolutions of the maps used in this study. In the left panels, the shaded area indicates the spatial scale larger than the FoV size of the CO observations, corresponding to the vertical dotted lines on the right panels. The y-axis shows the range of possible correlation coefficients, from -1.0 (negative correlation) to +1.0 (positive correlation), with the horizontal dashed line indicating no correlation (coefficient value = 0). }
    \label{fig:cc}
\end{figure*}

Narrow \Hone\ components are thought to represent clouds in the process of thermally condensing, in contrast to the broader \Hone\ components. At high physical resolutions, their spatial correlation with molecular gas clouds or regions of star formation may break down. In \citet{kim2023phangs} and in a review by \citet{Schinnerer2024molecular}, the authors investigated the morphological offsets between H$\alpha$ peaks, which trace recent star formation, and CO peaks in the spiral galaxy NGC~628. These studies followed similar work by \citet{schruba2010scale} and \citet{kruijssen2014uncertainty}. The authors found that the lifetime of a cloud, before it forms stars, leads to a spatial deviation between the H$\alpha$ and CO core peaks, reducing the morphological correlation between SFR and molecular gas mass—particularly when the physical resolution of the maps is smaller than 200-300~pc. In this context, narrow \Hone\ components are expected to have a longer lifetime before becoming unstable and cooler, making them more likely to eventually form stars. This suggests that a better spatial correlation with SFR may emerge at scales larger than 300~pc for narrow \Hone\ components.

To test the impact of spatial resolution on the relationship between narrow/broad \Hone, molecular gas (H$_{2}$), and SFR, we perform a cross-correlation analysis to identify the spatial scale ($k$) where the correlations are strongest. First, we compute the 2D power spectrum of each image, following the procedures outlined by \citet{martin2015ghigls}, \citet{blagrave2017ghigls}, and \citet{marchal2021resolving}. We apply apodization to all 2D maps used in the cross-correlation analysis before converting them to Fourier space to mitigate Gibbs ringing \citep[See Appendix A in ][]{mivilledeschenes2002power}. Specifically, sharp edges in the 2D maps can introduce the Gibbs phenomenon during the Fourier transform, leading to artificially elevated values along the axes in the Fourier plane. This effect potentially results in misinterpretation of the power spectrum. To reduce this, we use a cosine kernel with a rectangular boundary following \citet{blagrave2017ghigls} that smoothly tapers the edges (excluding the central 70\% of the image) along both the x- and y-axes.

For the apodized 2D image, $I(x, y)$, the power spectrum is the square of the magnitude of the Fourier-transformed intensity at spatial scale $k$:
\begin{equation}
    P(k_x, k_y) = |F(k_x, k_y)|^2,
\end{equation}
and the power spectrum $P(k_x, k_y)$ represents the intensity of different spatial frequency components in the image.

We then compute the correlation coefficient between the power spectra of the surface density maps of the narrow and broad \Hone\ components and those of molecular gas and SFR at different spatial scales ($k$). Since CO observations are typically limited to the central regions of galaxies, we crop the narrow and broad \Hone\ gas maps to match the FoV of the CO observations. Consequently, the cross-correlation analysis with molecular gas components is performed only within the central region. For SFR maps, the FoV is typically larger than CO observations, but it may still be smaller than that of the \Hone\ data. Therefore, to ensure a fair comparison of the power spectra, we crop the narrow and broad \Hone\ gas maps to align with the region where SFR estimates are available.

Fig.~\ref{fig:cc} shows the cross-correlation analysis for the narrow (green) and broad (orange) \Hone\ components of our sample galaxies, comparing them with the molecular surface density ($\Sigma_{\rm mol}$) in the left panels and the SFR surface density ($\Sigma_{\rm SFR}$) in the right panels. For most galaxies, the correlation coefficients between both narrow and broad \Hone\ components and $\Sigma_{\rm mol}$, as well as SFR, fluctuate between -0.5 and 0.5 across all spatial scales. This suggests a low or negligible correlation. As expected, the cross-correlation results for the narrow and broad \Hone\ components with molecular gas and SFR follow similar trends, reflecting that SFR is primarily regulated by molecular gas content (black lines in both panels). Below, we highlight some characteristic features of each galaxy based on our cross-correlation analysis.

First, for Sextans~A, the narrow \Hone\ components show a moderate correlation (coefficient value $> 0.5$) with $\Sigma_{\rm mol}$ and $\Sigma_{\rm SFR}$ at $\sim$0.7~kpc, corresponding to the size of the star-forming regions on the southeast and northwest sides of the galaxy (see Fig.~\ref{fig:maps_dwarf}). The broad \Hone\ components, on the other hand, show no particular trends, with low correlation coefficients across all scales. In NGC~6822, the narrow \Hone\ components show a positive correlation with $\Sigma_{\rm mol}$ and $\Sigma_{\rm SFR}$ at k $\sim$ 0.5--0.7~kpc, and k $\sim$ 0.7~kpc, respectively. Broad \Hone\ components have the strongest correlation with both $\Sigma_{\rm mol}$ and $\Sigma_{\rm SFR}$ at k $\sim$ 0.7. Similar to Sextans~A, in WLM, the narrow \Hone\ components exhibit a moderate correlation coefficient (r $=$ $\sim$0.5) at $k \sim 0.6$~kpc, corresponding to the sizes of the star-forming regions in the galaxy. The broad \Hone\ components, however, show the strongest correlation with SFR at $k \sim 0.8$~kpc, a larger scale than where the narrow \Hone\ components show the strongest correlation. 

The correlation coefficients between molecular gas and SFR in the flocculent spiral galaxies, NGC~5068 and NGC~7793, are much lower, ranging between 0.6 and 0.3, and even dropping to near 0 at smaller scales ($k < 1$~kpc) in NGC~7793. This may be due to the turbulence induced by the gravitational instabilities, which is a widely accepted explanation for the flocculent, unconnected arms of these galaxies \citep[][]{elmegreen2003turbulent}. In NGC~5068, the broad \Hone\ components are more strongly correlated with both molecular gas and SFR compared to the narrow \Hone\ components. A similar trend is observed in NGC~7793, where the broad \Hone\ components show a higher correlation with molecular gas for most spatial scales.

No noticeable features are observed in the cross-correlation analysis for NGC~1566, apart from that broad \Hone\ components have a better correlation with molecular gas and SFR at most scales. However, due to poor physical resolution for this galaxy (the smallest spatial scale is $\sim$1.1~kpc), and the lack of molecular gas information limited by the FoV, it is challenging to interpret the connection between different gas-phases.

For NGC~5236, the correlation coefficients between molecular gas and SFR are consistently near 1, down to a spatial scale of 1.5~kpc. Both narrow and broad \Hone\ components show no correlation with molecular gas or SFR for most scales, except at $k \sim$2.5~kpc, where broad \Hone\ components exhibit a moderate correlation with both molecular gas content and SFR. This spatial scale corresponds to the size of the galaxy’s spiral arms, which is also evident in the maps. 

In summary, although our cross-correlation analysis shows no clear trends between narrow \Hone\ components and molecular gas or SFR, we find that in dwarf galaxies, the narrow \Hone\ distribution is more strongly correlated at spatial scales of approximately 500–700~pc. In contrast, spiral galaxies do not exhibit a preferred spatial scale for stronger correlation.


 
\section{Discussion}
\label{sec:discussion}
\subsection{Radial trend of metallicity, SFR density, and \fn}
\label{sec:discussion-radialtrend}

\begin{figure*}
    \centering
    \includegraphics[width=0.98\linewidth]{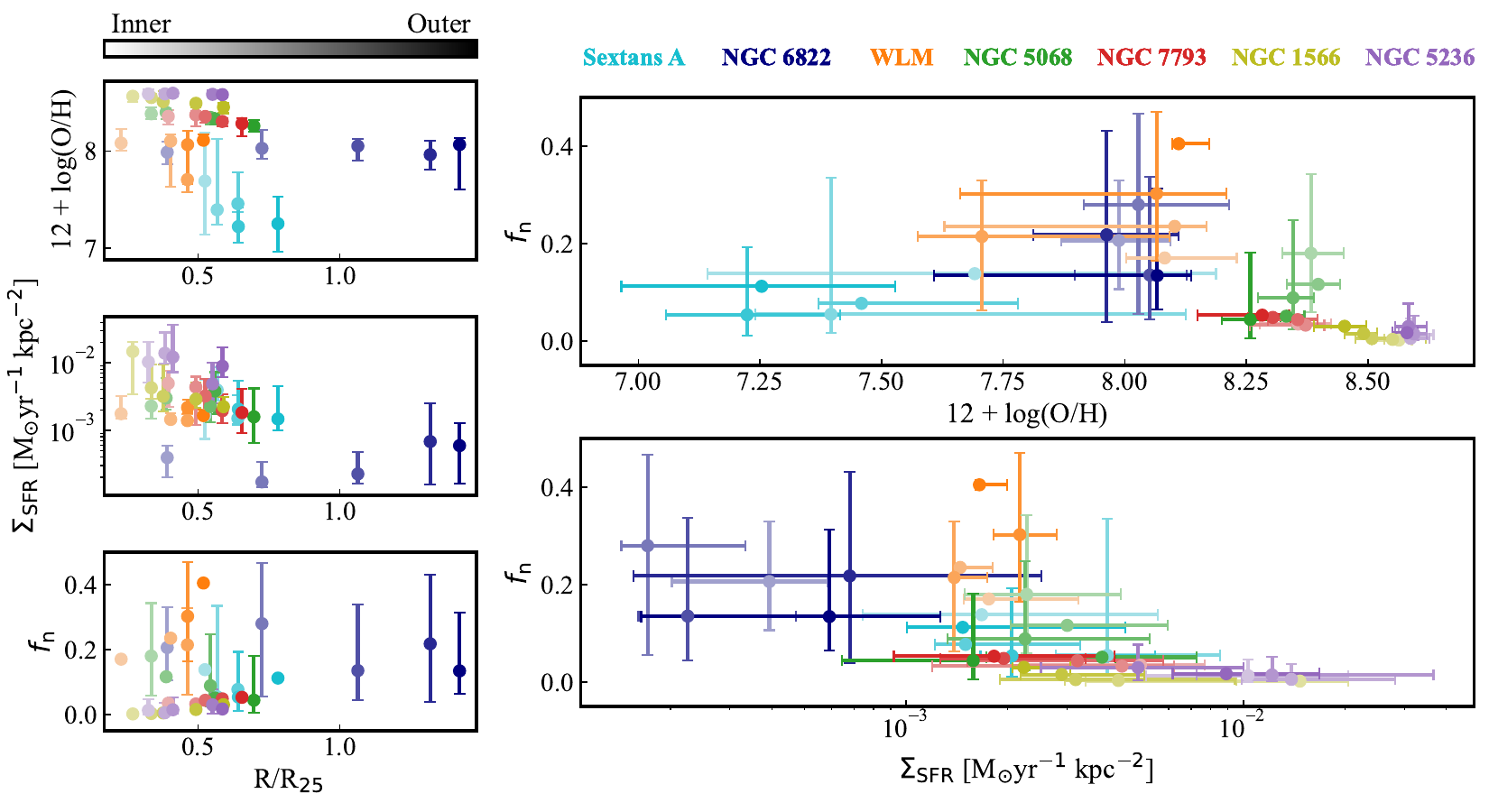}
    \caption{Left panels: Radial trends of gas-phase metallicity (\logOH), SFR surface density ($\Sigma_{\rm SFR}$), and \fn\ for galaxies. Each point represents the median of regions in each radial bin, with the bins defined by equal numbers of regions. The median measurement uncertainties and the 16\% and 84\% percentiles of the distribution within each radial bin are combined in quadrature and shown as error bars. Points are colour-coded for each galaxy, as indicated in the top right of the figure, with darker colours corresponding to the outer regions of the galaxy. Right panels: The trend of radially binned \fn\ as a function of radially binned gas-phase metallicity and SFR surface density.}
    \label{fig:radial_fn_sfr_logoh}
\end{figure*}

In this section, we investigate how local environmental factors, such as gas-phase metallicity (\logOH) and the SFR surface density, have an impact on the narrow \Hone\ fraction (\fn) by analysing the radial trend. In the three left panels of Fig.~\ref{fig:radial_fn_sfr_logoh}, we show the variation of gas-phase metallicity, the SFR surface density, and \fn\ as a function of radius, normalised by the optical radius of each galaxy. The \fn\ values are binned radially, with each bin containing an equal number of regions, and are compared to the \logOH\ and $\Sigma_{\rm SFR}$ in the corresponding radial bins, as shown in the two panels on the right. We take the median value and the corresponding 16\% and 84\% percentile of the values within the radial bin.

Taking advantage of the TYPHOON survey, which provides a wide FoV of optical emission line maps, we explore metal abundance up to larger radii in galaxies and briefly discuss the radial gradients observed in our sample. The dwarf galaxies—Sextans~A, NGC~6822, and WLM—show larger fluctuations in metallicity compared to spiral galaxies, and two of them (NGC~6822 and WLM) exhibit no clear positive or negative metallicity gradient. Sextans~A shows a negative gradient, however, it is likely dominated by the lower metallicity in the NW regions of the galaxy (Fig.~\ref{fig:maps_dwarf}). This lack of a gradient is consistent with the weak metallicity gradients found in dwarf galaxies from both cosmological baryonic zoom-in simulations and observations \citep[e.g., FIRE-2; ][]{mercado2021relationship, porter2022spatially}. This behaviour is attributed to the shallower gravitational well of dwarf galaxies, which allows metals injected from star formation to mix more evenly throughout the system \citep[][]{porter2022spatially}. For our spiral galaxy sample—NGC~5068, NGC~7793, NGC~1566, and NGC~5236—the metallicity gradients have been previously covered by \citet{grasha2022metallicity} as part of the same TYPHOON survey. All four spiral galaxies exhibit a negative gradient, in line with the results from \citet{grasha2022metallicity} and other references. This supports the concept of inside-out galactic growth \citep[][]{boissier1999chemo}.

The radial distribution of the SFR surface density (middle panel on the left) shows a decreasing trend with radius for all spiral galaxies, whereas the dwarf galaxies have a different pattern. For instance, Sextans~A and WLM display reduced star formation activity in their central regions (attributed to the \Hone\ supershell feature in the centre), followed by an increase in SFR density in the next radial bin, with a subsequent decline at the outer edges.

Regarding the \fn\ trend as a function of radius, dwarf galaxies show random fluctuations in \fn\ across different radii. In contrast, spiral galaxies show more defined radial trends. Except for NGC~5068 (green dots), which exhibits a negative radial gradient up to 0.7R/R$_{25}$ (corresponding to 3.8~kpc), the other spiral galaxies show higher \fn\ values at their outer regions.

On the right side of Fig.\ref{fig:radial_fn_sfr_logoh}, we show the relationships between gas-phase metallicity and \fn\ (upper panel) and between SFR surface density and \fn\ (lower panel), binned radially (inner regions - brighter, outer regions - darker). Dwarf galaxies show no clear trends of \fn\ with respect to gas-phase metallicity or SFR surface density. In contrast, for spiral galaxies, except for NGC~5068, \fn\ is generally higher in regions with low metallicity and low SFR surface density, which are typically located at larger galactocentric radii. Several factors may influence these trends: (1) the inside-out growth of spiral galaxies, where colder or narrow \Hone\ gas, which is thought to condense thermally from the diffuse warm neutral medium toward molecular gas, serves as a reservoir for star formation; (2) the transition of \Hone\ to molecular gas being more efficient in the inner regions of galaxies where metallicities are higher; and (3) stellar or AGN feedback, which can ionise neutral \Hone\ gas, broaden its line, and make it more difficult for the colder \Hone\ gas from to be observed.

\subsection{Limitations of this study}
\label{sec:discussion-limitations}
\begin{figure*}
    \centering
    \includegraphics[width=0.9\linewidth]{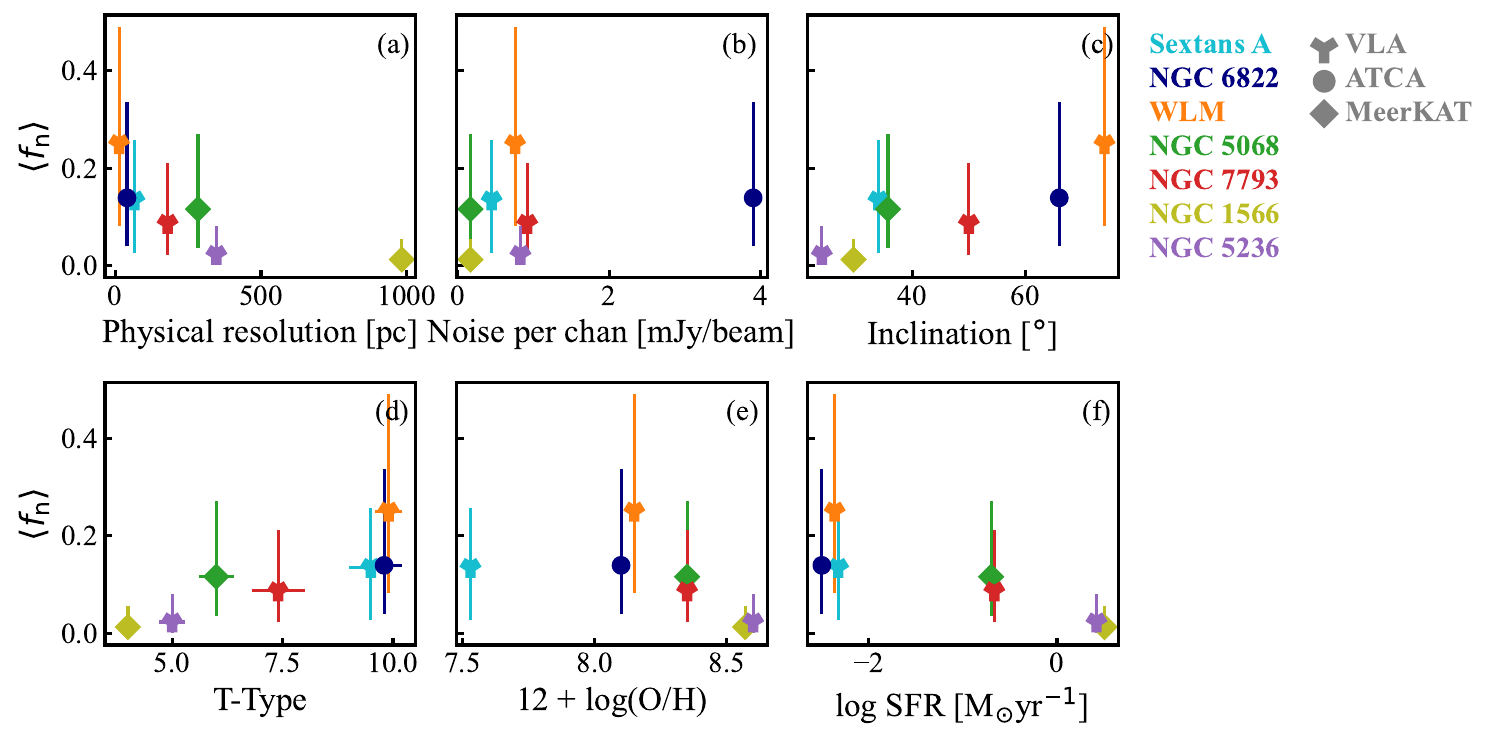}
    \caption{The median \fn\ and 16\% and 84\% percentiles for each galaxy as a function of the physical resolution of datasets (a), sensitivity in noise per channel (b), galaxy inclination (c), morphological T-Type (d), integrated gas-phase metallicity (e), and integrated SFR (f). Each galaxy is colour-coded as in Fig.~\ref{fig:radial_fn_sfr_logoh}, with distinct symbols representing each facility, as listed on the right side of the figure.}
    \label{fig:fn_res_incl_sfr_logoh}
\end{figure*}

In Fig.~\ref{fig:fn_res_incl_sfr_logoh}, we show the median value of \fn, \medianfn, with its 16\% and 84\% percentiles represented as error bars, across all star-forming regions in a galaxy (i.e., excluding pixels without SFR measurements). The upper panels display the trend of \medianfn\ in relation to observational limitations/factors, including the physical resolution of the datasets, the sensitivity of the original \Hone\ data, and the inclination of the galaxy. The bottom panels show how \medianfn\ correlates with galaxy properties such as morphological type, metallicity, and SFR. Data points are colour-coded according to Fig.~\ref{fig:radial_fn_sfr_logoh}, and different symbols are used for VLA, ATCA, and MeerKAT data, helping to examine how different observational strategies affect the \medianfn\ results (e.g., sensitivity, spectral resolution).

In panel (a) of Fig.~\ref{fig:fn_res_incl_sfr_logoh}, \medianfn\ shows a strong decreasing trend at coarser physical resolutions, which is expected. The \Hone\ velocity profiles reflect different ISM phases and kinematics, which become blended at lower resolutions. This suggests that higher physical resolution is necessary to accurately identify narrow components using Gaussian decomposition methods. In Appendix~\ref{app:physres_effect}, we test how physical resolution affects our ability to identify narrow components by degrading the spatial resolution of \Hone\ data cubes of three galaxies with a different range of inclination (Sextans~A, NGC~7793, and WLM) using \texttt{CASA-imsmooth} and \texttt{CASA-imrebin}, and running \baygaud\ on each resulting data cube. We find that \medianfn\ tends to decrease with decreasing resolution for the three galaxies tested. From 10,000 bootstrap samples of the median and scatter, the mean correlation coefficient between physical resolution and \medianfn\ is --0.29, --0.09, and --0.29, for Sextans~A (34$^\circ$), NGC~7793 (50$^\circ$), and WLM (74$^\circ$), respectively, indicating a low to moderate negative correlation.

The sensitivity of the \Hone\ 21~cm spectra can influence the ability to detect narrow components, particularly those associated with low column density \Hone\ components. In panel (b), we explore whether the sensitivity of the \Hone\ data cubes affects the narrow \Hone\ fraction. However, no correlation is found. Due to the variety of datasets from galaxies at different distances, it is inconclusive from the figure whether sensitivity influences our ability to identify colder \Hone\ components.

The detection of narrow \Hone\ components can also be affected by the inclination of a galaxy (panel c of Fig.~\ref{fig:fn_res_incl_sfr_logoh}). A higher fraction of narrow components is observed in edge-on galaxies (on the right-hand side), possibly because multiple gas clouds with different kinematics are aligned along the same line of sight. This can make the velocity profile non-Gaussian, facilitating the decomposition of individual clouds \citep[][]{oh2019robust}. On the other hand, this effect might be counteracted by increased line-of-sight blending, where multiple clouds contribute to the observed velocity profile. As a result, the impact of inclination remains inconclusive in this study.

In panels (d) (e), and (f) of Fig.~\ref{fig:fn_res_incl_sfr_logoh}, we observe decreasing trends in \medianfn\ for galaxies with a lower morphological T-Type \citep[][]{devaucouleurs1959classification}\footnote{Retrieved from HyperLeda, \url{http://atlas.obs-hp.fr/hyperleda/}}, higher metallicity, and higher SFR (i.e., more massive/structured spirals), respectively. These trends may reflect a more efficient \Hone\ to \Htwo\ transition in spiral galaxies, which have a larger amount of coolants (metals) and higher gas pressure compared to metallicity-poor dwarf galaxies, especially in their inner regions \citep[e.g., ][see also Fig.~\ref{fig:radial_fn_sfr_logoh}]{elmegreen1993h, honma1995molecular}. Spiral galaxies with high metallicity contain large amounts of molecular gas, which is concentrated in the central regions and spiral arms (see Fig.\ref{fig:maps_spirals}). Additionally, the low \medianfn\ observed in spiral galaxies could be attributed to the high SFR, which induces turbulence in the surrounding medium, broadening the \Hone\ 21~cm line profile \citep[][]{agertz2009large}. 

\subsection{The path forward: cold \Hone\ detection in external galaxies with future 21~cm surveys}
As discussed in Sec~\ref{sec:discussion-limitations}, spatial resolution (i.e., physical resolution) is a crucial factor in detecting colder \Hone\ components. Additionally, the sensitivity of the data plays a significant role in detecting \Hone\ gas components with low column densities. The upcoming \Hone\ facilities, such as the next-generation VLA (ngVLA) and the Square Kilometre Array (SKA), are expected to provide valuable information about colder \Hone\ components in external galaxies, offering much-improved sensitivity and spatial resolution compared to currently available \Hone\ facilities.

To effectively search for the cold \Hone\ gas phase, a finer channel resolution of $< 2$\kms\ is required, given that the expected velocity dispersion induced by thermal broadening is 1--2\kms. Among our sample galaxies, \Hone\ data for NGC~5068 and NGC~1566 come from one of the most recent \Hone\ deep surveys, MHONGOOSE \citep[][]{deblok2024mhongoose}, which utilises the SKA precursor, MeerKAT. The MHONGOOSE survey observes galaxies with unprecedented sensitivity (down to $< 10^{18}$ cm$^{-2}$ at 3$\sigma$ noise per channel), spatial resolution ($\leq 10$\arcsec), and spectral resolution ($\sim$ 1.4\kms), advancing our ability to study cold \Hone\ components, low column density \Hone\ gas (e.g., tidal interaction, gas accretion), and detailed gas kinematics \citep[][]{healy2024possible}. Additionally, the Local Group L-Band Survey, which observes six nearby galaxies \citep[][]{pingel2024local}, offers high spatial and spectral resolution ($\sim$0.42\kms), further enhancing our understanding of cold \Hone\ components in external galaxies, through techniques such as absorption lines (including \Hone\ self-absorption), Gaussian decomposition, and more. More rigorous analyses of \Hone\ kinematic decomposition will be possible with the completion of these programs and facilities.

\section{Conclusion}
\label{sec:summary}
In this study, we analyse the spatial distribution and radial trends of colder, narrow \Hone\ gas components in nearby galaxies, examining their relationship with gas-phase metallicity, molecular gas, and star formation rate to understand how these factors influence the presence and the fraction of the colder \Hone\ gas. Our sample is selected from the TYPHOON IFS survey, which provides gas-phase metallicity measurements out to large galactocentric distances (larger than the optical radius). We compile \Hone\ data of the selected sample galaxies, from the THINGS (VLA), LITTLE THINGS (VLA), MHONGOOSE (MeerKAT), and individual ATCA observations. We identify colder, narrow \Hone\ components by performing an optimal Gaussian decomposition of \Hone\ 21~cm data cubes for seven nearby galaxies, spanning a wide range of metallicities.

Using the kinematic decomposition tool \baygaud, we fit velocity profiles with up to three or four components and classify them as narrow (colder) or broad \Hone\ based on a velocity dispersion threshold of $< 6$\kms. We analyse the spatial distribution of narrow and broad components and their morphological association with molecular gas (from CO) and star formation (from FUV+MIR). We examine the radial trends of the narrow \Hone\ fraction (\fn), gas-phase metallicity, and SFR surface density, as well as their relationships in radially binned values. Finally, we compare the median \medianfn\ of each galaxy with observational limitations (physical resolution, sensitivity, and inclination) and galaxy properties (morphological type, integrated metallicity, and integrated SFR). We summarise our main findings below.

\begin{enumerate}
    \item Multiple \Hone\ kinematic components are preferentially found in star-forming regions and spiral arms of galaxies, whereas diffuse gas is more commonly observed with a single \Hone\ component.
    \item The ability to perform kinematic decomposition depends on spectral resolution and sensitivity. Galaxies from the deep \Hone\ MHONGOOSE survey (NGC~5068 and NGC~1566), observed at high spectral resolution ($\sim$1.4\kms), tend to exhibit more complex velocity profiles, often requiring more Gaussian components.
    \item The mass fraction of the colder \Hone\ components, \fn, defined by the 6\kms\ velocity dispersion threshold and bulk motioning, are sub-dominate to the broad \Hone\ components and range from 3\% to 16\% across our sample galaxies.
    \item The colder, or narrow, \Hone\ components show a clumpier and more filamentary distribution compared to the broad \Hone\ components.
    \item Dwarf galaxies have a higher (stronger) correlation between narrow and molecular gas or SFR at around 500--700~pc from cross-correlation analysis, whereas no clear correlation at a specific spatial scale is found for spiral galaxies.
    \item Radially binned \fn\ shows no clear correlation with gas-phase metallicity or SFR for dwarf galaxies. Spiral galaxies exhibit a decreasing trend of \fn\ with higher metallicity and higher SFR, likely driven by inside-out growth, \Hone-to-\Htwo\ transition in the inner regions, and the influence of stellar or AGN feedback. 
    \item The physical resolution plays a crucial role in identifying narrow \Hone\ components, as evidenced by the lower median \medianfn\ in galaxies with coarser dataset resolution.
    \item Even considering the physical resolution of the data, the \medianfn\ is higher for dwarf galaxies, having high morphological T-Type value, low metallicity, and low SFR, compared to spiral galaxies. This implies 1) the efficient transition from \Hone-to-\Htwo\ gas in spiral galaxies, derived by the larger amount of coolants (high metallicity) and high gas pressure, and 2) the velocity profile broadening of neutral \Hone\ gas by higher SFR and AGN in the centre region, and the subsequent turbulence in spiral galaxies. 
\end{enumerate}

\section*{Acknowledgements}
The authors thank the anonymous referee who has provided constructive and helpful comments to improve the paper. H-JP thanks Nickolas Pingel for providing background source information for the NGC~6822 absorption line detections. H-JP also appreciates Se-Heon Oh for the insightful discussions. BFM thanks The Observatories of the Carnegie Institution for Science for believing in and supporting this decades-long programme at the duPont telescope at Las Campanas, Chile. KG is supported by the Australian Research Council through the Discovery Early Career Researcher Award (DECRA) Fellowship (project number DE220100766) funded by the Australian Government. Parts of this work are supported by the Australian Research Council Centre of Excellence for All Sky Astrophysics in 3 Dimensions (ASTRO~3D), through project number CE170100013.

This research has made use of the NASA/IPAC Extragalactic Database, which is funded by the National Aeronautics and Space Administration and operated by the California Institute of Technology. This publication makes use of data products from the Wide-field Infrared Survey Explorer, which is a joint project of the University of California, Los Angeles, and the Jet Propulsion Laboratory/California Institute of Technology, and NEOWISE, which is a project of the Jet Propulsion Laboratory/California Institute of Technology. WISE and NEOWISE are funded by the National Aeronautics and Space Administration. The MeerKAT telescope is operated by the South African Radio Astronomy Observatory, which is a facility of the National Research Foundation, an agency of the Department of Science and Innovation. The Australia Telescope Compact Array is part of the Australia Telescope National Facility (\url{https://ror.org/05qajvd42}) which is funded by the Australian Government for operation as a National Facility managed by CSIRO. We are grateful to the National Radio Astronomy Observatory (NRAO) for time on the VLA and the people at the VLA who make things run. The VLA is a facility of the NRAO, itself a facility of the National Science Foundation that is operated by Associated Universities, Inc. We are also grateful to the National Science Foundation for funding the LITTLE THINGS project with grants to Hunter (AST-0707563), Elmegreen (AST-0707426), Simpson (AST-0707468), and Young (AST-0707835) over the period June 2007-June 2012. This work made use of THINGS `The HI Nearby Galaxy Survey \citet[][]{walter2008things}'. The Joint ALMA Observatory is operated by ESO, AUI/NRAO and NAOJ. The National Radio Astronomy Observatory is a facility of the National Science Foundation operated under a cooperative agreement by Associated Universities, Inc. This study makes use of the following ALMA data: 
ADS/JAO.ALMA\#2013.1.01161.S,
ADS/JAO.ALMA\#2015.1.00121.S,
ADS/JAO.ALMA\#2015.1.00925.S,
ADS/JAO.ALMA\#2015.1.00956.S,
ADS/JAO.ALMA\#2016.1.00386.S,
ADS/JAO.ALMA\#2017.1.00392.S,
ADS/JAO.ALMA\#2017.1.00766.S,
ADS/JAO.ALMA\#2017.1.00886.L,
ADS/JAO.ALMA\#2018.1.00484.S,
ADS/JAO.ALMA\#2018.1.01321.S,
ADS/JAO.ALMA\#2018.1.01651.S,
ADS/JAO.ALMA\#2018.A.00062.S,
ADS/JAO.ALMA\#2019.1.01235.S,
ADS/JAO.ALMA\#2018.1.00337.S, 
ADS/JAO.ALMA\#2019.2.00110.S, 
ADS/JAO.ALMA\#2021.1.00330.S. 
ALMA is a partnership of ESO (representing its member states), NSF (USA) and NINS (Japan), together with NRC (Canada), NSTC and ASIAA (Taiwan), and KASI (Republic of Korea), in cooperation with the Republic of Chile. This work has made use of data from the European Space Agency (ESA) mission {\it Gaia} (\url{https://www.cosmos.esa.int/gaia}), processed by the {\it Gaia} Data Processing and Analysis Consortium (DPAC, \url{https://www.cosmos.esa.int/web/gaia/dpac/consortium}). Funding for the DPAC has been provided by national institutions, in particular the institutions participating in the {\it Gaia} Multilateral Agreement. We acknowledge the usage of the HyperLeda database (\url{http://leda.univ-lyon1.fr}).

This work made use of Astropy:\footnote{http://www.astropy.org} a community-developed core Python package and an ecosystem of tools and resources for astronomy \citep{astropy2013, astropy2018, astropy2022}. This work also used Numpy (\citealt{numpy}), Matplotlib (\citealt{matplotlib}), and Scipy (\citealt{scipy}).

This research was conducted on Ngunnawal Indigenous land.

\section*{Data Availability}

The TYPHOON optical IFS datacubes will be made publicly available in a forthcoming release by Seibert et al. (in prep.) through Data Central\footnote{\url{https://datacentral.org.au/}}. The emission line data products can be made available upon reasonable request by emailing A. Battisti.




\bibliographystyle{mnras}
\bibliography{bib} 




\appendix

\section{Physical resolution effect on the retrieving narrow \Hone\ components}
\label{app:physres_effect}

\begin{figure*}
    \centering
    \includegraphics[width=0.64\linewidth]{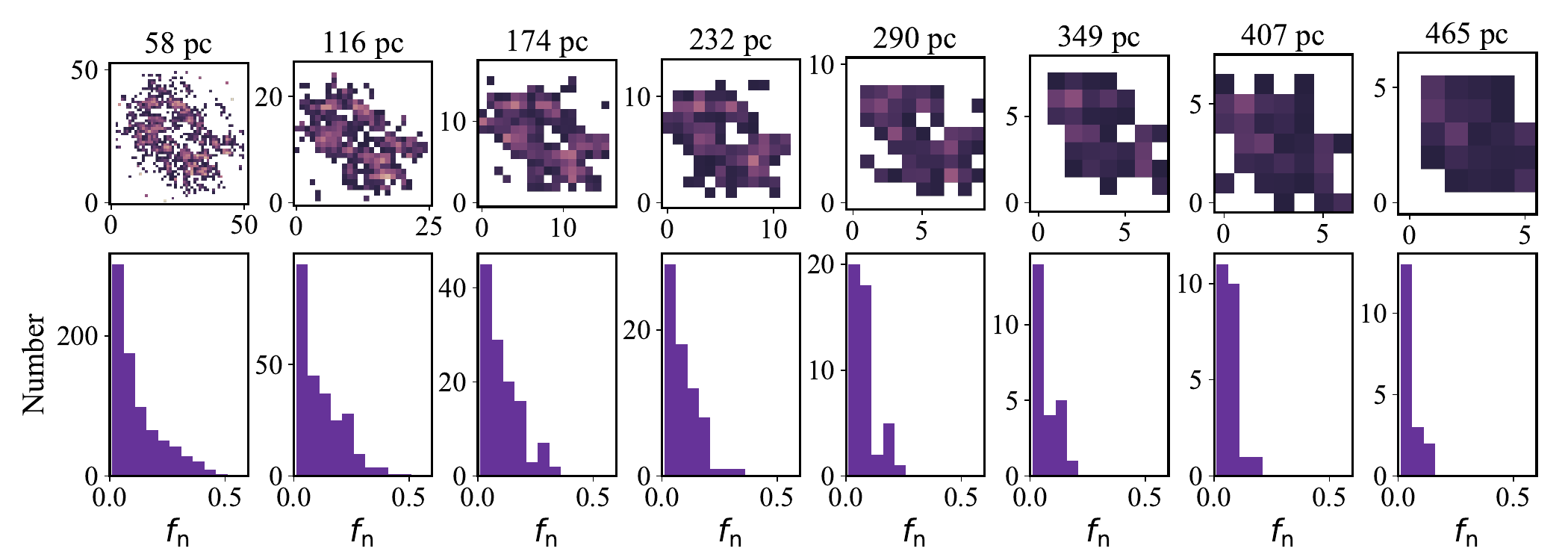}
    \includegraphics[width=0.35\linewidth]{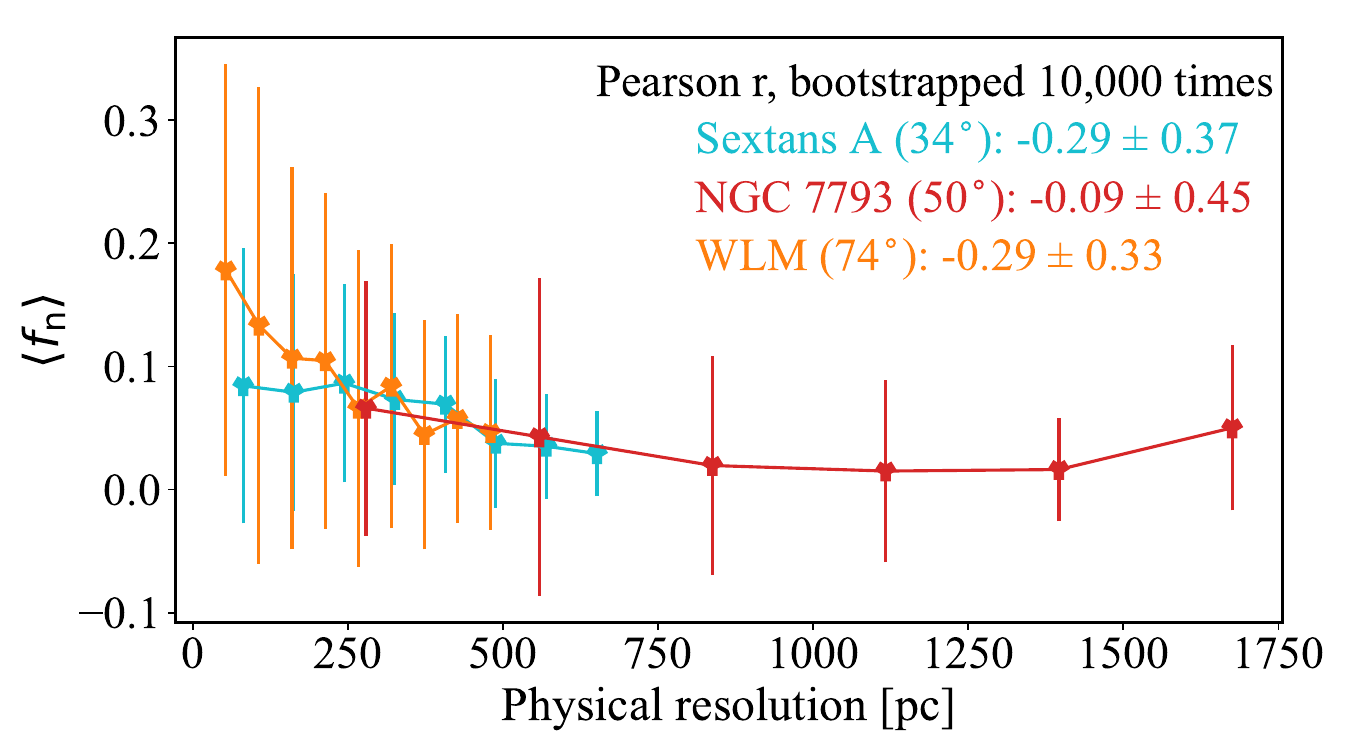}
    \caption{Left: \fn\ maps (top) of Sextans~A at each physical resolution and the histogram (bottom) as an example. Right: the tendency of the median \fn\ and the 68\% confidence interval as error bars as a function of physical resolution for Sextans~A (cyan), NGC~7793 (red), and WLM (orange).}
    \label{fig:sextansa_fn_tendency_resolution}
\end{figure*}

Fig.~\ref{fig:sextansa_fn_tendency_resolution} shows the maps of \fn\ of Sextans~A at different physical resolutions (top panels), its histogram (bottom panels) on the left as an example and the median \fn\ as a function of the physical resolution for three galaxies with different inclination (Sextans~A, NGC~7793, and WLM) on the right. The vertical error bars indicate the 68\% confidence interval) of \fn\ at each physical resolution. The right panel indicates that \medianfn\ tends to decrease with a coarser resolution. We test this with the bootstrapping method with a Gaussian distribution of the median and the standard deviation of \medianfn\ and calculate the Pearson correlation coefficient $r$. With 10,000-resample steps, we find that the mean $r$ is --0.29, --0.09, and --0.29 for Sextans~A, NGC~7793, and WLM, respectively, with a large error of $>$ 0.3, implying a low to moderate anti-correlation \citep[see also Section 4.1 in][]{ianjasimanana2012shapes}. However, it still implies that the physical resolution of the \Hone\ data can impact the recovery of narrow or colder \Hone\ components, suggesting that higher resolution data are needed to determine the ideal spatial scale where the value of \fn\ converges.

\section{Galaxy properties}

\textit{Sextans~A} ($Z\sim$~0.1~\Zsol) is a dwarf irregular (dIrr) galaxy and the lowest-metallicity system in our sample. The galaxy hosts three well-known, distinct star-forming regions \citep[][]{vandyk1998recent, dohmpalmer2002deep}, along with the relatively recent discovery of Region~D by \citet{garcia2019ongoing} (see their Fig.~1). \citet{garcia2019ongoing} also found a spatial correlation between the \Hone\ distribution and OB stars, suggesting that neutral gas may fuel star-forming activities in this low-metallicity system.

\textit{NGC~6822} ($Z\sim$~0.3~\Zsol) is the closest external dIrr galaxy after the Magellanic Clouds, located at a distance of $\sim$~0.5~Mpc. The star-forming regions \citep[e.g.,][]{karampelas2009star} and molecular clouds \citep[e.g.,][]{schruba2017physical, park2024spatially} are concentrated in the central region of the galaxy, which exhibits a bar-like structure with a position angle (PA) of 10$\degree$ \citep[][]{mateo1998dwarf}. In contrast, the neutral gas traced by the \Hone\ 21cm line is far more extended, covering an area of about 1$\degree$ on the sky with a PA of 118$\degree$ \citep[][]{deblok2006star}. A study of the colder \Hone\ components in NGC~6822 by \citet{park2022gas} \citep[see also][]{deblok2006star} revealed a weak but noticeable correlation between the surface densities of the cool \Hone\ components (classified by a velocity dispersion threshold of 4\kms) and the SFR.

\textit{WLM} ($Z\sim$~0.13~\Zsol), or Wolf-Lundmark-Melotte, is an isolated, low-metallicity dIrr galaxy. Despite its low metallicity, \citet{rubio2015dense} reported the discovery of 10 small molecular gas clouds in the galaxy, detected for the first time in such a low-metallicity system using ALMA CO observations. Subsequent ALMA observations with broader coverage by \citet{archer2022environments} revealed that regions of high \Hone\ surface density are not always associated with molecular gas cores. However, molecular gas cores consistently exhibit high \Hone\ surface densities, indicating a possible threshold for molecular gas formation in such environments from the neutral phase.

\textit{NGC~5068} ($Z\sim$~0.6~\Zsol) is a barred spiral galaxy. Recent MeerKAT \Hone\ 21~cm observations of this galaxy \citep[][]{healy2024possible}, conducted as part of the MHONGOOSE survey \citep[][]{deblok2024mhongoose}, showed anomalous \Hone\ features in velocity profiles. These features are characterised by a high-velocity dispersion of $\sim$~18\kms, significantly higher than the typical value expected for star-forming clouds. Notably, it is located outside the galaxy's optical radius on the northwestern side, suggesting a possible origin related to gas accretion.

\textit{NGC~7793} ($Z\sim$~0.6~\Zsol) is a flocculent spiral galaxy. Its distinct distribution of star-forming clouds and ISM is thought to result from its stochastic star formation history. Using Very Large Array (VLA) data from the THINGS survey \citep[][]{walter2008things}, \citet{saikia2020gas} explored the two-component \Hone\ gas observed along the line of sight throughout the galaxy, identifying a bimodal distribution in velocity dispersion.

\textit{NGC~1566} ($Z\sim$~0.9~\Zsol) is the most metal-rich and the most distant galaxy in our sample at distance $\sim$~18~Mpc. It has an integrated SFR of 4.468~\Msolyr\ \citep[][]{leroy2019z}, primarily concentrated in its spiral arms. The galaxy exhibits an asymmetric \Hone\ disk with a slightly warped structure, which is thought to result from ram pressure stripping due to interactions with the intergalactic medium \citep[][]{elagali2019wallaby}. Another possibility is that the past interaction with the nearby galaxy, NGC~1581, as shown in \citet[][]{maccagni2024mhongoose}.

\textit{NGC~5236} ($Z\sim$~0.9~\Zsol), also known as M83, is a barred spiral galaxy featuring prominent spiral arms and an active galactic nucleus (AGN). It has a high SFR of 4.167~\Msolyr\ \citep[][]{leroy2019z}, concentrated primarily in the central region. Observations of the \Hone\ 21~cm line using VLA \citep[][]{walter2008things} show a deficiency of neutral gas in the central region, which is attributed to factors such as the complete conversion of \Hone\ to \Htwo, beam smearing, or stellar feedback driving neutral gas outward. NGC~5236 also hosts an extended UV disk beyond $\sim 2r_{\rm 25}$ \citep[][]{thilker2005recent}, indicative of a prolonged star formation history over 1~Gyr \citep[][]{dong2008spitzer}. The XUV disk is found to correlate well with \Hone\ \citep[][]{bigiel2010tightly}.

\bsp	
\label{lastpage}
\end{document}